\documentclass[hidelinks]{article}
\usepackage{arxiv}
\usepackage[utf8]{inputenc} 
\usepackage[T1]{fontenc}    
\usepackage{url}            
\usepackage{booktabs}       
\usepackage{amsfonts}       
\usepackage{nicefrac}       
\usepackage{microtype}      
\usepackage{graphicx}
\usepackage[square,sort,comma,numbers]{natbib}
\usepackage{doi}
\usepackage{graphicx}
\usepackage{subfigure}
\usepackage{miller}
\usepackage{amsmath}
\usepackage{amssymb}
\usepackage{chemformula}
\usepackage{float}
\usepackage{color,soul}
\usepackage{mathtools}
\usepackage{appendix}
\usepackage{color}
\usepackage{tikz}
\usepackage{siunitx}
\usepackage{lineno}
\usepackage[normalem]{ulem}
\usepackage{upgreek}
\usepackage{siunitx}
\usepackage{fancyhdr}

\graphicspath{{figures/}}

\title{Sensitivity of grain-averaged elastic strain and orientation predictions on the mesh density and boundary conditions in crystal plasticity finite element simulations}

\author{
    Jeremiah Lethoba \\
	Department of Mechanical Engineering\\
	The University of Alabama\\
	Tuscaloosa AL 35487 \\
	\texttt{jlethoba@crimson.ua.edu} \\
	    \And
	Romain Quey \\
    Mines Saint-Etienne \\
    Univ Lyon, CNRS \\
    UMR 5307 LGF \\
    F-42023 Saint-Etienne, France \\
	\texttt{romain.quey@cnrs.fr} \\
        \And
	Darren C. Pagan \\
    Department of Materials Science and Engineering \\
    The Pennsylvania State University \\
    University Park, PA 16802 \\
	\texttt{dcp5303@psu.edu} \\
        \And
	Matthew Kasemer \\
	Department of Mechanical Engineering\\
	The University of Alabama\\
	Tuscaloosa, AL 35487 \\
	\texttt{mkasemer@eng.ua.edu} \\
}

\renewcommand{\shorttitle}{Lethoba et. al. 2026}

\hypersetup{
pdftitle={Sensitivity of grain-averaged elastic strain and orientation predictions on the mesh density and boundary conditions in crystal plasticity finite element simulations},
pdfsubject={cond-mat.mtrl-sci},
pdfauthor={Jeremiah Lethoba, Romain Quey, Darren C. Pagan, Matthew Kasemer},
pdfkeywords={Crystal plasticity, Finite element analysis, Mesh sensitivity, Grain-averaging, High energy X-ray diffraction microscopy},
}

\begin{document}
\maketitle

\begin{abstract}
    Combined high-energy X-ray diffraction microscopy (HEDM) and crystal plasticity finite element (CPFE) modeling studies have emerged as a preferred paradigm to shed insight into the evolution of elasticity and plasticity at the intragrain scale of polycrystals. In particular, far-field HEDM measures the deformation response of upwards of thousands of individual grains simultaneously {\emph{in situ}} during mechanical loading, though measurements are primarily limited, however, to the {\emph{average}} state of each grain---i.e., the grain's full strain tensor, crystallographic orientation, spatial location and volume. CPFE is utilized to shed information on the {\emph{intragrain}} deformation response, via the sub-discretization of each grain into many finite elements, though the direct point of comparison to HEDM remains the grain-averaged response. We thus seek to find the minimum simulation conditions necessary to provide consistent grain-averaged predictions in an attempt to limit computational cost. In this study, we perform a suite of simulations and systematically study the effects of mesh density and boundary conditions, and consider different materials. We discuss these results and show that accurate prediction of grain-averaged elastic strains in a given region of interest typically requires a mesh with 250 elements per grain on average and a buffer layer of at least three grains between the region of interest and the control surfaces.
\end{abstract}


\section{Introduction}
\label{sec:introduction}

Since at least the early 1930s, the field of micromechanics with application to polycrystalline materials has been active in attempting to relate observations regarding the evolution of elasticity and plasticity at lower length scales to the macroscopic response~\cite{kockstomewenk}. In the ensuing decades, the relative use and spatial fidelity of various experimental techniques to both characterize microstructures and observe or measure the deformation response has increased. These methods include the use of electron microscopy (in coordination with image processing methods, such as digital image correlation~\cite{Echlin2016}), micropillar mechanical testing~\cite{Kodam2023}, micro- and nanoidentation~\cite{Mukhopadhyay2006}, and others to provide quantitative information regarding how crystals and ensembles of crystals deform under mechanical loading. However, these micromechanical methods are often limited to the measurement or observation at or near the surfaces of samples. Consequently, this limitation often renders results with decreased confidence of bulk-representative behavior.

Most recently, a wide array of non-destructive, three dimensional X-ray and neutron techniques have been developed to experimentally study the grain scale deformation response of polycrystalline aggregates. The most mature of these techniques is far-field high-energy X-ray diffraction microscopy~\cite{margulies2002strain,poulsen2004three} (ff-HEDM, or alternatively 3DXRD). In ff-HEDM, X-ray scattering (i.e., measured diffraction peaks) is directly associated with individual grains. This enables the statistically significant measurement of the deformation response of up to thousands of grains simultaneously at the microscale, {\em in situ} during deformation loading~\cite{miller2020understanding}. These measurements have primarily included face-centered cubic~\cite{margulies2002strain,juul2017measured,tayon2019situ,tayon2024exploring}, body-centered cubic \cite{wang2017direct,ball2023three,Mengiste2024}, and hexagonal close-packed~\cite{oddershede2011grain,wang2017direct,peterson2025determining} alloy systems, but has been extended to other crystalline systems including geological materials~\cite{hurley2018situ}, concrete~\cite{nair2019micromechanical,hurley2019situ}, and solid-state electrolytes \cite{dixit2022polymorphism}. While these diffraction peaks indeed contain information about intragrain behavior~\cite{pagan2014connecting,naragani2017investigation,Renversade2024}, with current detector technology, ff-HEDM is best adapted to track the average motion of these peaks, from which we may reconstruct the \textit{grain-averaged} crystallographic orientation and full elastic strain tensor, as well as the centroidal position and volume of the grain~\cite{bernier2011far,nygren2020algorithm}. From the grain-averaged elastic strain tensors, the full stress state in individual grains can then be calculated using the anisotropic form of Hooke's law.

In tandem with experimental advances, deformation modeling of polycrystalline materials has likewise focused increasingly on the influence of phenomena at lower length scales. The work of Schmid and Boas~\cite{Schmid1935} formalized the elastic and inelastic mechanics of single crystals, while the contemporary development of the mean-field approaches of Sachs~\cite{Sachs1928} and Taylor~\cite{Taylor1934a,Taylor1934b,Taylor1938} focused on the extension to the consideration of ensembles of grains, though with no care to consider grain-to-grain interaction. This lead to the later development of viscoplastic self-consistent modeling~\cite{Molinari1987,Lebensohn1994}, which introduces a global form of grain interaction (via an effective medium), and local grain-to-grain interaction models~\cite{VanHoutte2005ALAMEL,Engler2005GIA,Quey2010RSI}. The last three decades have witnessed the advent and refinement of full-field deformation prediction frameworks, in particular crystal plasticity finite element (CPFE) modeling~\cite{Marin_1998,Marin_1998b,Roters_2010}. CPFE considers high fidelity representations of large ensembles of grains (i.e. polycrystals)~\cite{Quey2011}, where each grain can be discretized into upwards of thousands of finite elements, allowing for the prediction of inhomogeneous intragrain deformation fields via the consideration of crystal-scale elasticity and plasticity  models~\cite{Kasemer2017,Kasemer2017b,Kasemer2020,Kasemer2020b,Cappola2021}.

Overall, HEDM data provide a natural complement to micromechanical modeling such as CPFE, facilitating representative sample instantiation~\cite{Shankar2024}, as well as model development, calibration, and validation~\cite{Mengiste2024,Peterson2025}. However, the intragrain fields predicted via CPFE are at scales below what HEDM generally offers. In order to make one-to-one comparisons to HEDM, the predicted element-scale data must be averaged over individual grains. A question that naturally arises during this data reduction is: what mesh resolution is necessary in order for the grain-averaged predictions to converge to ensure proper comparisons with experiment? Stated another way, if the grain-averaged results are of primary interest, what is the least computational expense (i.e., smallest mesh) necessary to produce consistent results? Another recurring question is: how far from the imposed boundary conditions are the grain-averaged predictions affected?  In other terms, how large of a ``buffer layer'' must be considered between the region of interest and the control surfaces to avoid significant bias? The concept that finite element predictions may be sensitive to changes in the crystal~\cite{Chatterjee2018} or mesh statistics---particularly refinement--- or boundary conditions is not itself novel. In particular, many crystal plasticity studies include some nominal study of mesh sensitivity, though are generally focused on the convergence of macroscopic behavior or global average behavior of element-scale results, or sensitivity of predictions to element types~\cite{Choi2012,Keshavarz2013,Khadyko2016,Lim2019,Feather2020}, and study is often limited to a mere formality compared to the overall focus of the paper. There exists relatively limited inspection of the sensitivity of grain-averaged behavior to changes in mesh refinement~\cite{Lv2011,Feather2020}, as would be beneficial for understanding limitations with comparison to HEDM data, though existing studies do not perform in-depth analyses on the grain-averaged behavior.

Consequently, we here address this gap by systematically assessing the sensitivity of grain-averaged results to the degree of mesh refinement and to the distance to the boundary conditions. We first perform a suite of CPFE simulations on a fixed polycrystalline sample with varying mesh density (average number of elements per grain). We analyze the evolution of the grain-averaged results of the elastic strain state and the orientation (i.e., the typical results gathered from HEDM experimentation) as a function of the mesh density. We inspect results at various points through the deformation history (elastic regime, transition regime, and plastic regime) to understand deviations in predictions due to deformation history. We further extend the analysis by considering the worst-case grain as well as the effects of free surfaces. We then investigate the effect of the boundary conditions, before finally generalizing the analysis by varying the degree of elastic anisotropy.

\section{Crystal Plasticity Finite Element Modeling}
\label{sec:cpfem}

Here, we utilize the software package FEPX for deformation modeling~\cite{fepx,neperfepx,fepxweb}. FEPX is a non-linear finite element solver programmed specifically to model the anisotropic elastic and plastic deformation response of polycrystalline materials. Generally, FEPX solves for the deformation field on high-fidelity representations of polycrystalline samples generated via tessellations or directly from three dimensional characterization, where each grain may be discretized into (potentially) thousands of elements. FEPX is parallelized via OpenMPI routines, which facilitate cluster computing to allow for the simulation of very large meshes on the order of millions of degrees of freedom (i.e., allowing for the consideration of either many grains, high mesh density, or both). The formulation utilized in FEPX is well established~\cite{Marin_1998,Marin_1998b,Roters_2010}, and we choose to forego a detailed discussion of the kinematics and finite element implementation for sake of brevity, and instead here describe the central pieces of the models governing the elastic and plastic behavior of the material. Further information can be found in refs.~\cite{fepx,neperfepx,fepxweb}.

To begin, we utilize the anisotropic form of Hooke's law to model the elastic response of the material, or:
\begin{equation}
    \label{eq:hooke}
    \boldsymbol{\tau} = \mathbb{C}\left(\mathbf{q}\right)\boldsymbol{\varepsilon}^\text{e} \quad ,
\end{equation}
where the Kirchhoff stress tensor, $\boldsymbol{\tau}$, is linearly related to the elastic strain tensor, $\boldsymbol{\varepsilon}$, via the anisotropic stiffness tensor, $\mathbb{C}$, itself a function of the orientation of the crystal, parameterized here via a quaternion, $\mathbf{q}$. We note that the stiffness tensor is reduced due to major and minor symmetry, as well as the consideration of the crystal symmetry~\cite{nye}.

Regarding plasticity, we consider a rate-dependent restricted slip model to describe slip kinetics, following the form:
\begin{equation}
    \label{eq:slip_kinetics}
    \dot{\gamma}^k = \dot{\gamma}_0\frac{\tau^k}{\tau_c}\left|\frac{\tau^k}{\tau_c}\right|^{\frac{1}{m}-1} \quad ,
\end{equation}
where $\dot{\gamma}$ is the rate of shear on a slip system $k$, $\tau$ is the resolved shear stress and $\tau_c$ the current critical resolved shear stress, and $m$ is the rate dependence exponent. Each slip system may contribute to a unique simple shearing mode on the crystal, as governed by the crystal symmetry and our choice of slip systems. We consider the slip systems commonly observed at room temperature: the \hkl{1 1 1}\hkl<1 1 0> type slip systems for FCC crystals, \hkl{1 1 0}\hkl<1 1 1> type slip systems for BCC crystals, and the \hkl{0 0 0 1}\hkl<1 1 -2 0> (basal), \hkl{1 0 -1 0}\hkl<1 1 -2 0> (prismatic), and \hkl{1 0 -1 1}\hkl<1 1 -2 -3> (pyramidal) type slip systems for HCP crystals.

As plasticity develops, we allow the critical resolved shear stress to evolve as a function of the local development of plasticity. We model this hardening behavior via a saturation style model, or:
\begin{equation}
    \label{eq:hardening}
    \dot{\tau}_c = h_0 \left(\frac{\tau_s - \tau_c}{\tau_s - \tau_0}\right)\dot{\Gamma} \quad ,
\end{equation}
where $h_0$ is the fixed-state hardening rate, $\tau_s$ is the saturation value and $\tau_0$ the initial value of the critical resolved stress, and $\dot{\Gamma}$ is the sum of the shear rates of all slip systems at a local point.

We further allow for the orientation to evolve as a function of the local development of plasticity. We consider this via the relationship:
\begin{equation}
    \label{eq:reorientation}
    \dot{\boldsymbol{q}} = \frac{1}{2}\,\boldsymbol{\omega_q}\,\boldsymbol{q} \quad,
\end{equation}
where $\boldsymbol{\omega_q} = \left(0,\,\boldsymbol{\omega}\right)$ is a pure quaternion comprised of $\boldsymbol{\omega}$, or the vector form of the lattice reorientation rate computed from the plastic spin rate (itself a function of the local degree of plastic deformation) and the skew part of the plastic velocity gradient (see ref.~\cite{fepx}).

\section{Simulation Suite}
\label{sec:simulations}

To determine the effects of mesh density on the prediction of grain-averaged quantities, we perform a suite of simulations on a polycrystal discretized with a varying mesh density. In this section, we describe the generation of the simulations for this purpose.

\subsection{Tessellation and Mesh Generation}
\label{sec:tess_and_mesh}

Regarding the generation of the tessellation for use in simulations, we begin by noting the three guiding principles. First, we aim to generate a polycrystal with a domain morphology and a sufficient number of grains to both limit boundary effects as well as provide sufficient grain and orientation statistics. Second, we aim to reduce the effect of anomalous microstructural features (i.e., fine facets, elongated grains, wide grain size or shape distributions). Finally, we aim to generate a polycrystalline sample representative of a typical HEDM configuration, both in terms of physical length (of the gauge section) and grain statistics.

\begin{figure}[htb!]
    \centering
    \includegraphics[height=0.46\textwidth]{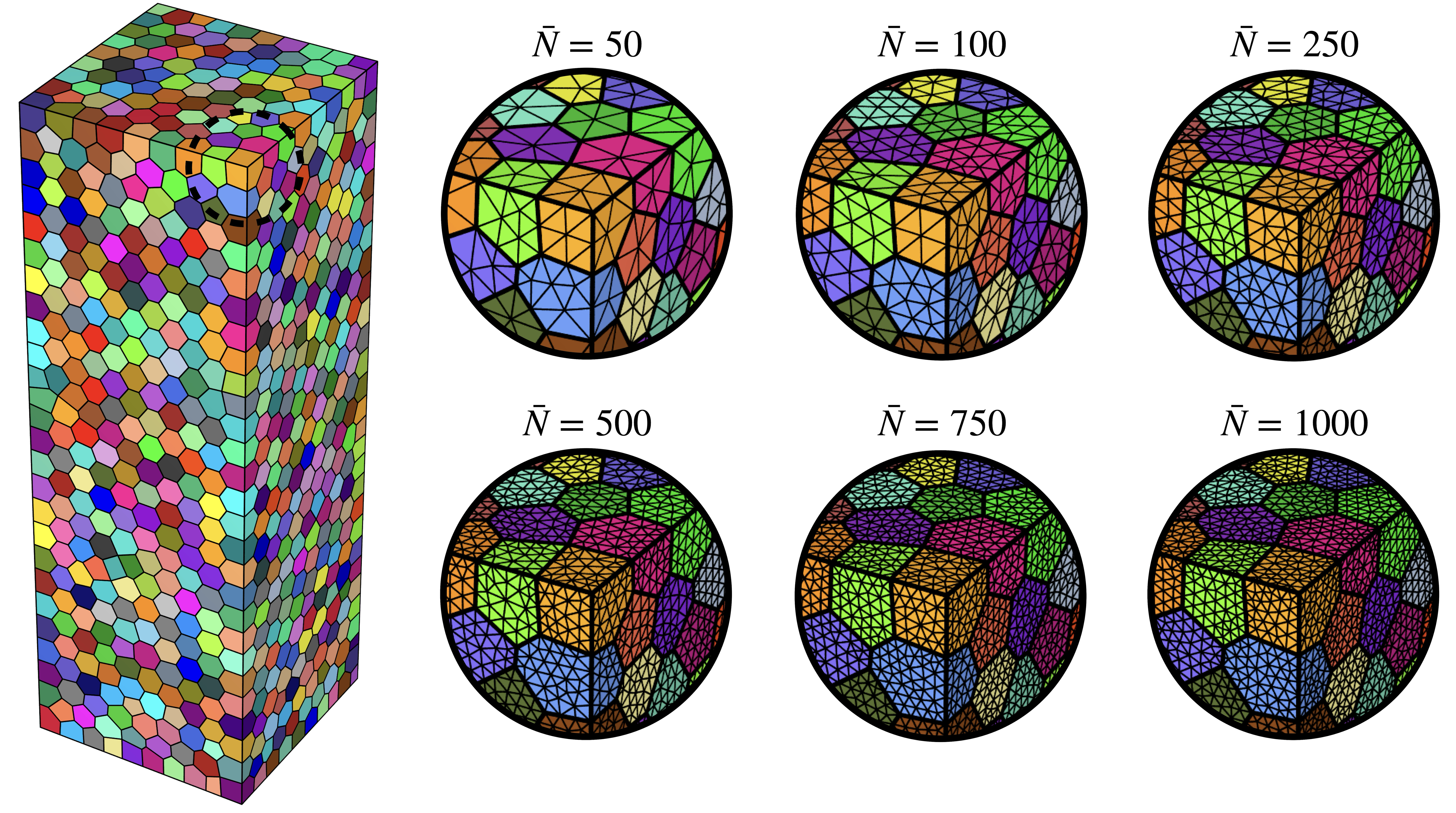}
    \caption{Visualizations of the tessellation representing the polycrystal utilized in the simulations with grains colored arbitrarily. Depictions of the meshes for a subset of the domain (indicated by a dashed black line on the tessellation) are shown for the six different mesh densities considered.}
    \label{fig:tess_and_mesh}
\end{figure}

Consequently, we generate a polycrystal of 3000 grains in a rectangular prismatic domain of dimension \qtyproduct[product-units = single]{1 x 1 x 3}{\milli\metre\cubed}.\footnote{We are careful to note here that the crystal plasticity model in this study considers no inherent length scale. We include units here for sake of familiarity and ease of discussion, rather than to imply absolutes.} This will allow us to consider only the central \qtyproduct[product-units = single]{1 x 1 x 1}{\milli\metre\cubed} region for the analysis, which contains about 1000~grains and provides sufficient statistics. Per Saint-Venant's principle, the control surfaces are sufficiently far from the region (10 grain sizes) that their effect can be considered as negligible. We utilize centroidal Voronoi tessellations to generate a highly regularized microstructure, which facilitates a mesh where each grain has roughly the same number of elements, and elements of the same size, reducing undue bias due to selective over- or under-refinement as would be expected from a morphology with a larger spread in grain size. We utilize the software package Neper~\cite{Quey2011,neperfepx,neperweb} to generate the tessellation, and present a visualization in Figure~\ref{fig:tess_and_mesh}. Upon generation of the tessellation, we assign orientations to each crystal within the volume. To provide robust orientation statistics, we choose a nearly uniform orientation distribution, so that every orientation (among all possible orientations) is equally represented~\cite{quey_uniform}.  We then assign the orientations randomly to the grains.

We next turn to mesh generation. Given the tessellation describing the geometric morphology of the polycrystal, we again utilize Neper to generate a geometry-conforming finite element mesh. To assess the effects of mesh density on the results, we hold the tessellation geometry and grain orientations fixed across all meshes. To parameterize the mesh density, we choose to target approximate values of the average number of elements per grain, denoted by $\bar{N}$. To achieve this, we manually optimize the relative characteristic length of the elements, $l^\text{rel}$, during mesh generation to achieve a total number of elements, $N$, to reach an approximate value for our target densities of $\bar{N}$ while maintaining a uniform and robust mesh. We summarize our mesh generation parameters in Table~\ref{tab:mesh_parms}, which further contains a summary of the mesh statistics for the five mesh densities we consider in this study. We present visualizations of the various mesh densities considered in this study in Figure~\ref{fig:tess_and_mesh}. As we will demonstrate in the following sections, the mesh with the highest density (i.e., $\bar{N}=1000$) presents a sufficiently high resolution to render converged results, and will thus serve as our point of reference against which other simulations will be compared. We further note that the mesh with the lowest density (i.e., $\bar{N}=50$) is at-or-near to the bottom density limit that we find achievable for grains of somewhat arbitrary shape, and thus represents the lower limiting case.
\begin{table}[H]
    \centering
    \caption{Mesh generation parameters and resulting mesh statistics for the various mesh densities considered in this study.}
    \begin{tabular}{ccc}
        ${\sim}\bar{N}$ (-) & $l^\text{rel}$ (-) & $N$ (-)  \\
        \hline
        50 & 1.5 & 151,997 \\
        100 & 1 & 323, 399 \\
        250 & 0.73 & 839,323 \\
        500 & 0.56 & 1,558,815 \\
        750 & 0.475 & 2,265,848 \\
        1000 & 0.44 & 2,775,681\\
    \end{tabular}
    \label{tab:mesh_parms}
\end{table}

\subsection{Material Parameters}
\label{sec:material_parameters}

Regarding material behavior, we choose to model the deformation response of Okegawa mold copper (OMC)~\cite{wong_obstalecki}. OMC displays a relatively high amount of elastic anisotropy, as well as significant hardening behavior. Further, OMC copper exhibits face-centered cubic symmetry, representative of a large class of materials. Collectively, we note that these aspects make OMC an interesting case to study. The elastic and plastic parameters that we choose for this study are summarized in Tables~\ref{tab:elastic_parms} and~\ref{tab:plastic_parms}, respectively, and are informed by ref.~\cite{wong_obstalecki}.
\begin{table}[H]
    \centering
    \caption{Single crystal elastic constants for OMC copper.}
    \begin{tabular}{cccc}
        $C_{11}$ (\SI{}{\giga\pascal}) & $C_{12}$ (\SI{}{\giga\pascal}) & $C_{44}$ (\SI{}{\giga\pascal}) \\
        \hline
        168.4 & 121.4 & 75.4 \\
    \end{tabular}
    \label{tab:elastic_parms}
\end{table}
\begin{table}[H]
    \centering
    \caption{Plasticity modeling parameters for OMC copper.}
    \begin{tabular}{ccccccc}
        $m$ & $\dot{\gamma}_0$ (\SI{}{\per\second}) & $h_0$ (\SI{}{\mega\pascal}) & $\tau_0$ (\SI{}{\mega\pascal}) & $\tau_{s}$ (\SI{}{\mega\pascal}) \\
        \hline
        0.02 & 1.0 & 400.0 & 17.0 & 122.4 \\
    \end{tabular}
    \label{tab:plastic_parms}
\end{table}

\subsection{Boundary Conditions and Loading History}
\label{sec:bcs_and_loading_history}

\begin{figure}[htb!]
    \centering
    \includegraphics[width = 2.5in]{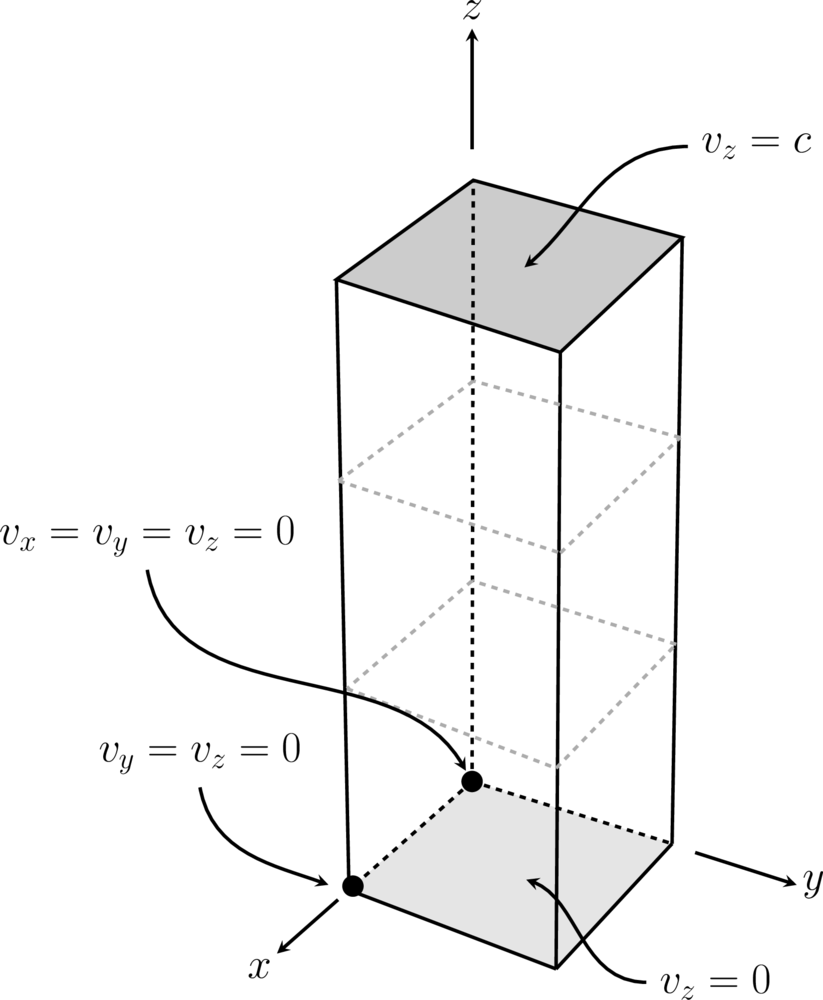}
    \caption{Boundary conditions utilized in simulations, in which uniaxial tension is applied via the imposition of velocities on the nodes belonging to two opposing surfaces of the domain (i.e., the control surfaces, where one surface is fixed and the other has a constant velocity, $c$, related to the macroscopic strain rate via the gauge length), and where two nodes are further restrained on the fixed control surface to arrest rigid body translation and rotation. Gray dashed lines further indicate the central region of the sample (Section~\ref{sec:tess_and_mesh}).}
    \label{fig:minimal_bc}
\end{figure}

We impose uniaxial deformation on the polycrystal in simulations via boundary conditions---visualized and summarized in Figure~\ref{fig:minimal_bc}---which apply uniaxial tension via the imposition of velocities on two opposing surfaces of the domain while otherwise restraining a minimal number of nodes to arrest rigid body translation and rotation. We set velocities on the nodes of the non-fixed extension control surface to enforce a macroscopic strain rate of \SI{1e-2}{\per\second}, chosen such that the deformation response is in the quasi-static loading regime. Regarding the loading history, we discretize time such that the imposed loading history provides data to \SI{10}{\percent} macroscopic strain, with enough temporal fidelity during the elastic-plastic transition to properly capture the yield behavior of the material. The same boundary conditions and loading profile are utilized for every simulation.

\section{Results}
\label{sec:results}

\begin{figure}[htb!]
    \centering

    \subfigure[]{
        \label{subfig:stress_strain_1}
        \includegraphics[width=2in]{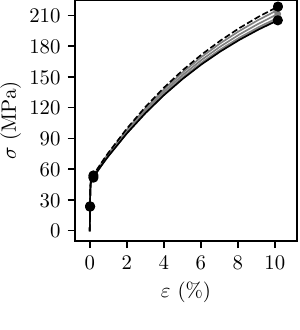}}
    \subfigure[]{
        \label{subfig:stress_strain_2}
        {\includegraphics[width=2.3in]{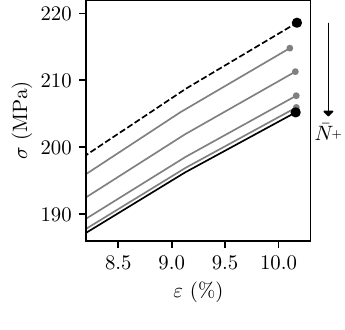}}}
    \caption{\subref{subfig:stress_strain_1} Macroscopic equivalent stress-strain behavior for the simulations performed with mesh densities of $\bar{N}=50$ (dashed line), $\bar{N}=1000$ (solid line), and the intermediary mesh densities (gray lines), and~\subref{subfig:stress_strain_2} detail of the behavior near \SI{10}{\percent} macroscopic strain.}
    \label{fig:stress_strain}
\end{figure}

Herein, we consider the results for the grains which are fully contained in the central region of the sample (described previously in Section~\ref{sec:tess_and_mesh}), which corresponds to 912~grains. We first plot the macroscopic stress-strain behavior in Figure~\ref{fig:stress_strain} to highlight the changes in the macroscopic behavior between the simulations performed with mesh densities of $\bar{N}=50$ and $\bar{N}=1000$. We highlight three areas of interest: a point in the elastic deformation regime (at a macroscopic strain of ${\sim}\SI{0.015}{\percent}$), near the point of yield (${\sim}\SI{0.2}{\percent}$), and in the fully developed plastic deformation regime (${\sim}\SI{10}{\percent}$). These macroscopic strain levels are representative of what is commonly probed in HEDM experiments.

\begin{figure}[htb!]
    \centering
    \subfigure[]{
        \label{subfig:mesh_strain_1}
        \includegraphics[height=2.60in]{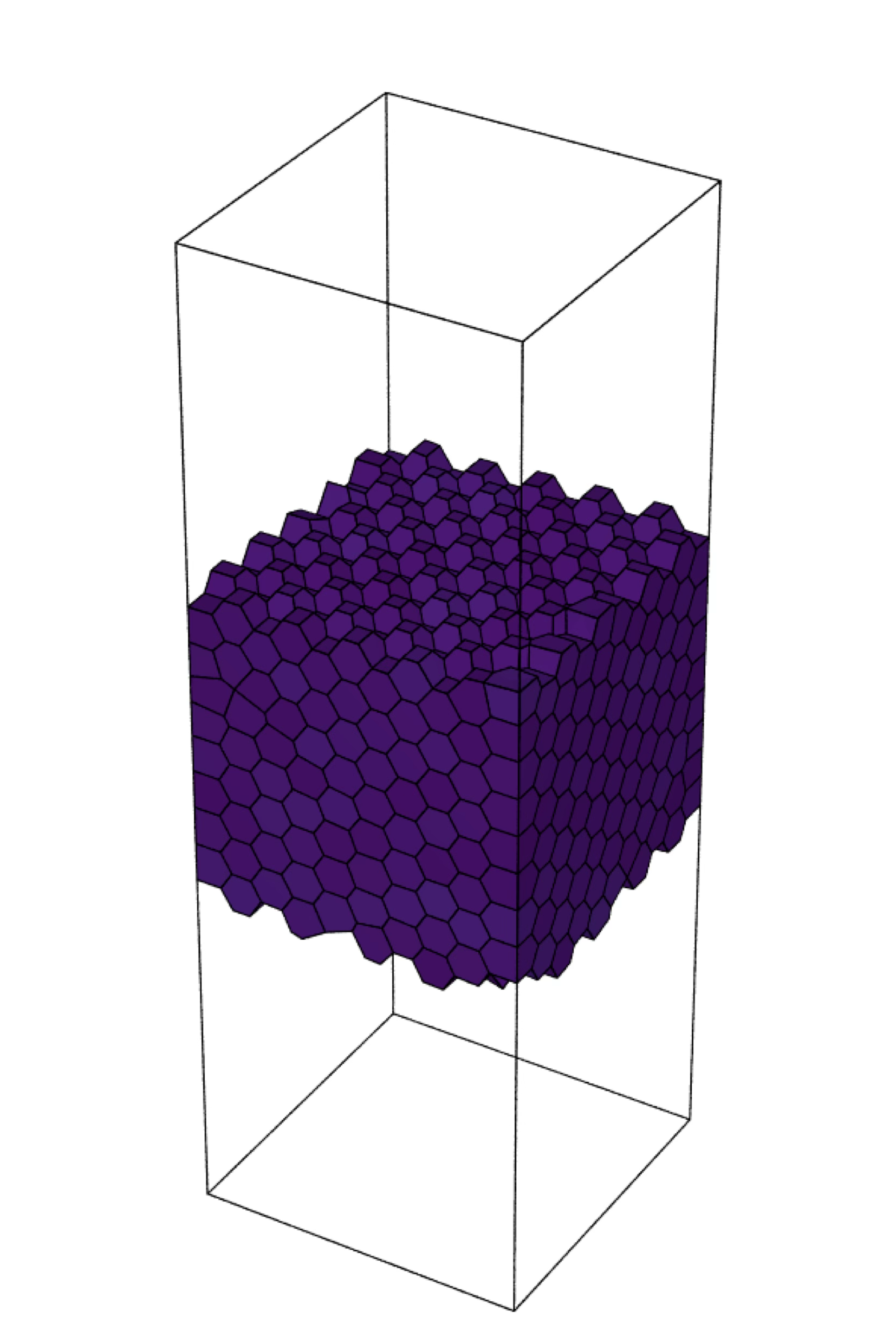}}
    \subfigure[]{
        \label{subfig:mesh_strain_2}
        \includegraphics[height=2.60in]{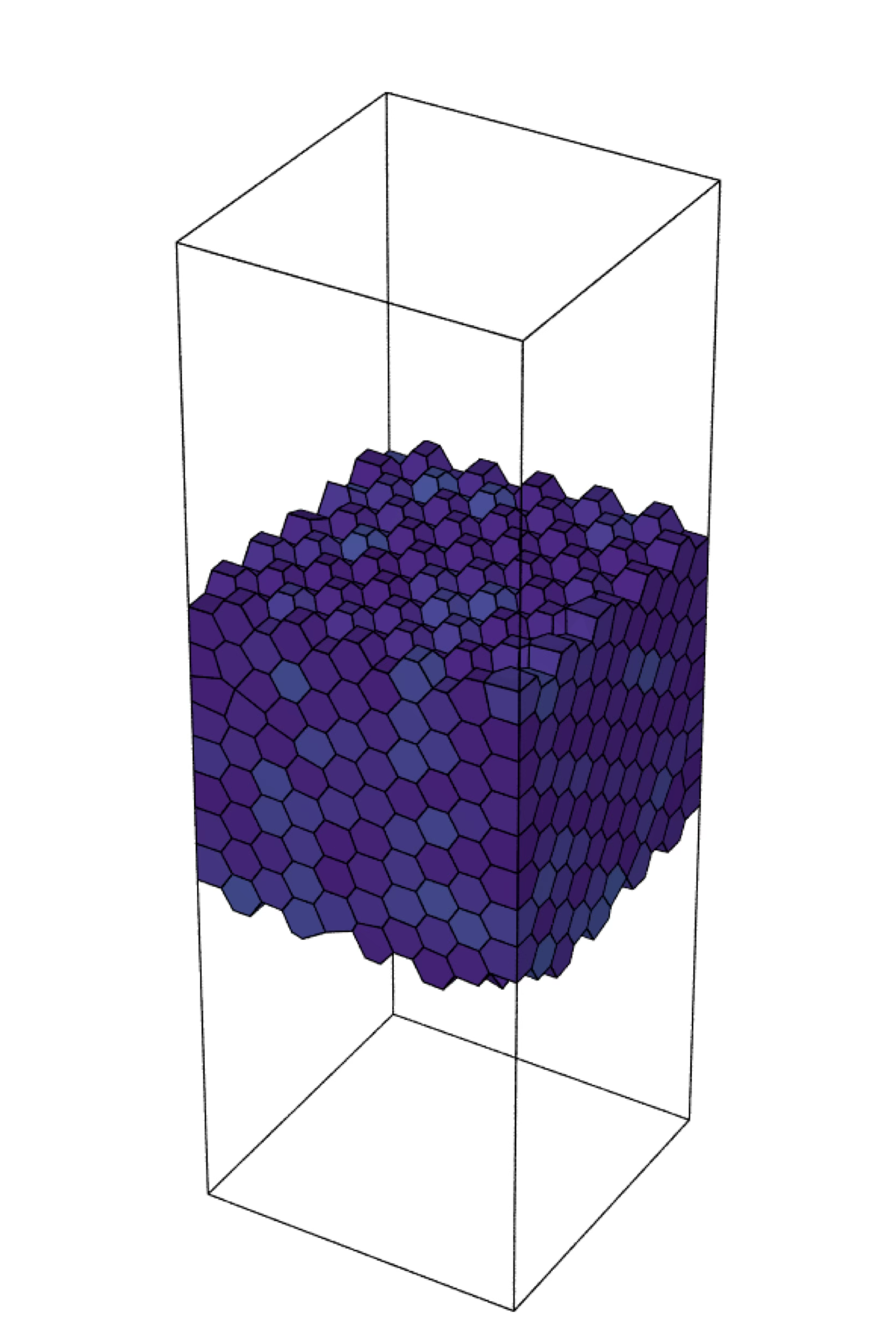}}
    \subfigure[]{
        \label{subfig:mesh_strain_3}
        \includegraphics[height=2.60in]{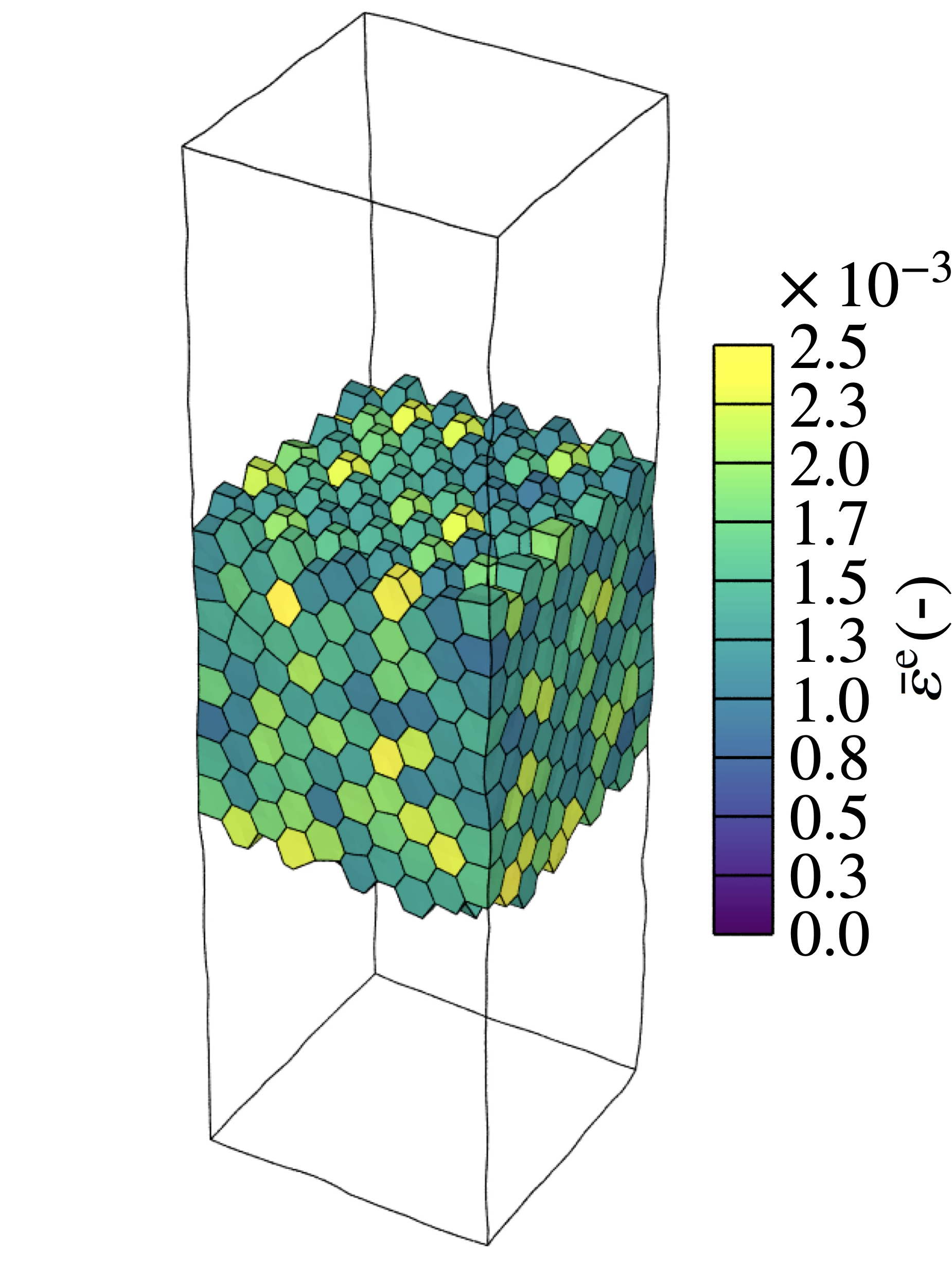}}
    \subfigure[]{
        \label{subfig:mesh_strain_4} 
        \includegraphics[height=2.60in]{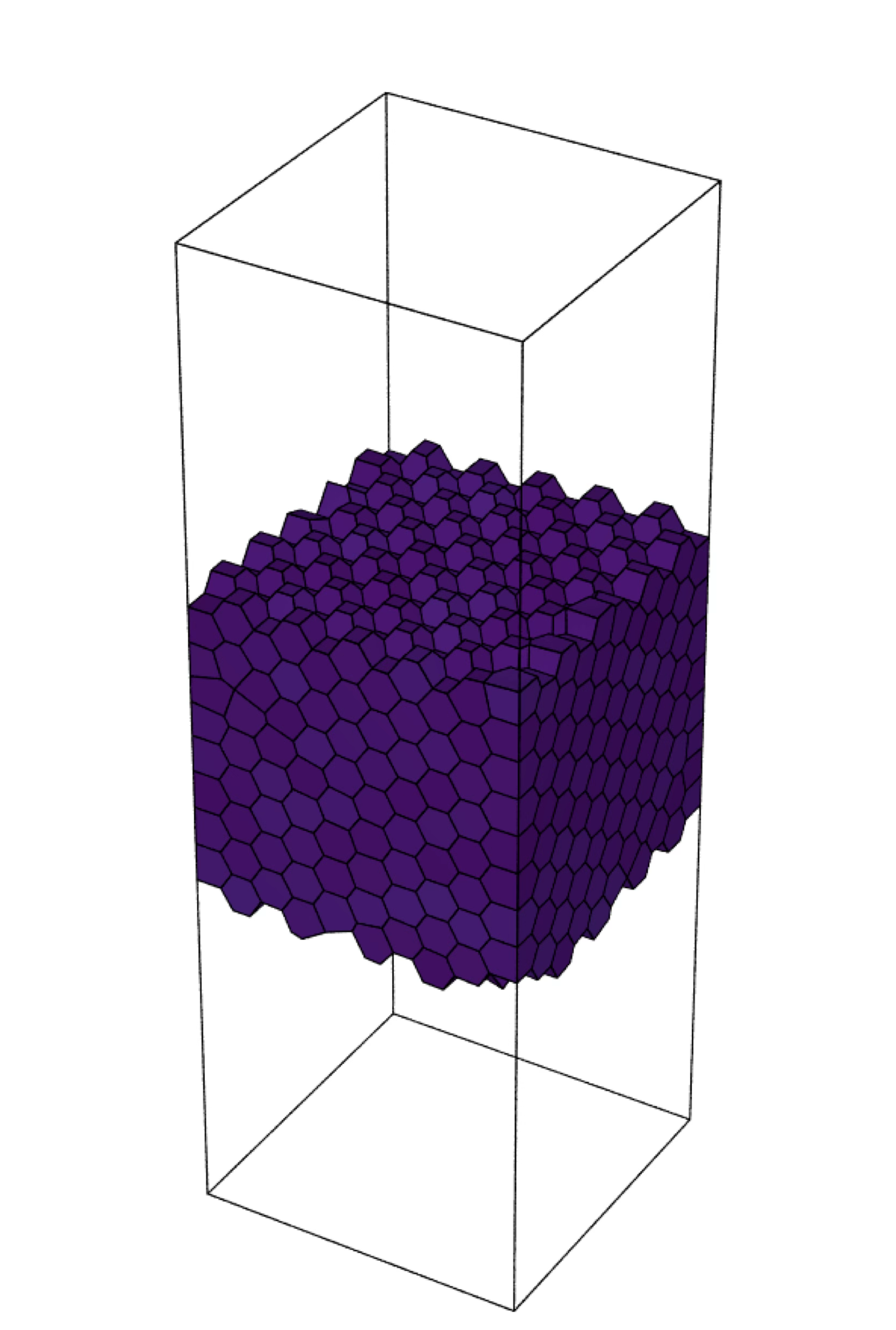}}
     \subfigure[]{
        \label{subfig:mesh_strain_5}
        \includegraphics[height=2.60in]{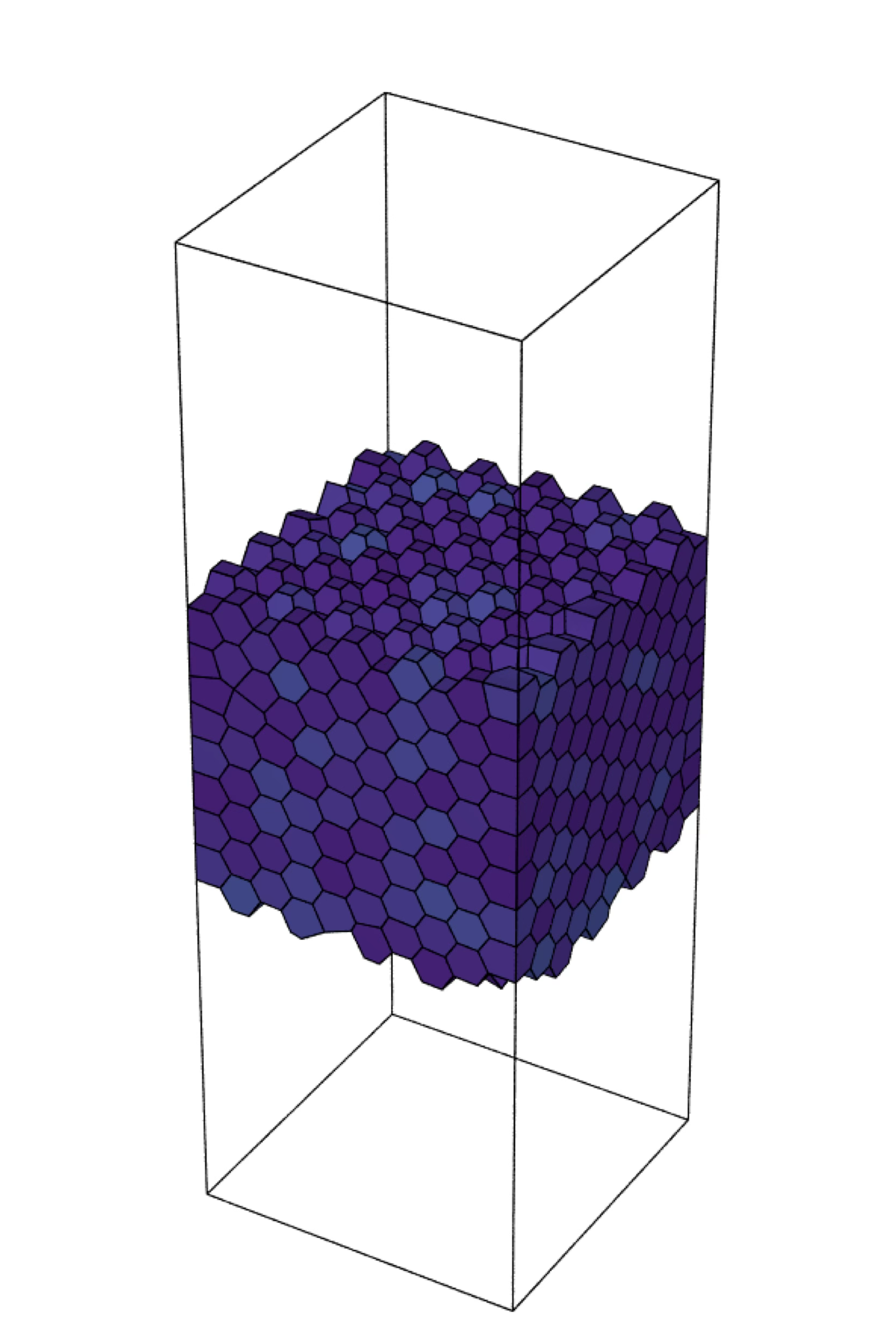}}
    \subfigure[]{%
        \label{subfig:mesh_strain_6}
        \includegraphics[height=2.60in]{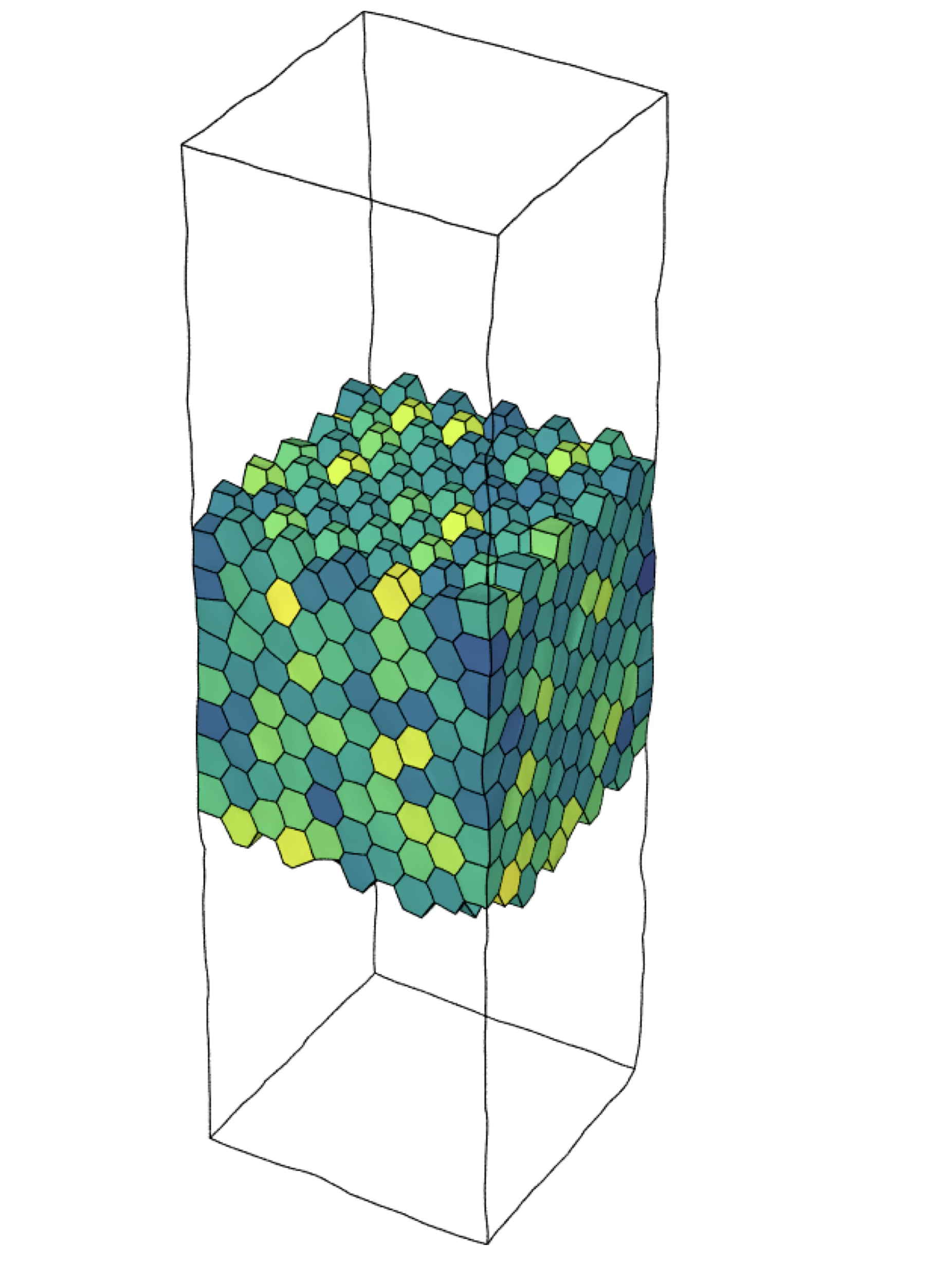}}
    \caption{Grain-averaged equivalent elastic strain, $\bar{\varepsilon}^\text{e}$, plotted spatially on the deformed meshes. Results are shown for the simulation with mesh density of $\bar{N}=50$ at macroscopic strains of \subref{subfig:mesh_strain_1} \SI{0.015}{\percent}, \subref{subfig:mesh_strain_2} \SI{0.2}{\percent}, and \subref{subfig:mesh_strain_3} \SI{10}{\percent} (including the common scale), and for the simulation with mesh density of $\bar{N}=1000$ at macroscopic strains of \subref{subfig:mesh_strain_4} \SI{0.015}{\percent}, \subref{subfig:mesh_strain_5} \SI{0.2}{\percent}, and \subref{subfig:mesh_strain_6} \SI{10}{\percent}.
}
    \label{fig:mesh_strain}
\end{figure}

We next plot the grain-averaged equivalent elastic strain states on the undeformed finite element mesh to qualitatively demonstrate the spatially inhomogeneous strain distributions across the domain of the polycrystal. We present this data in Figure~\ref{fig:mesh_strain}.

To systematically quantify the differences in strain predictions between the simulations, we formulate a grain-averaged strain error metric, $e$ (or $e_i$, for the $i$-th grain). This metric accounts for the differences between the grain-averaged equivalent elastic strain states (i.e., similar to what can be measured via HEDM) predicted in a simulation with a given mesh density to its counterpart in the simulation with the highest mesh density (i.e., our reference case of $\bar{N}=1000$, as described in Section~\ref{sec:tess_and_mesh}). We describe the formulation in the following.

\begin{figure}[htb!]
    \centering
    \subfigure[]{%
        \label{subfig:strain_error_1}
        \includegraphics[width=2in]{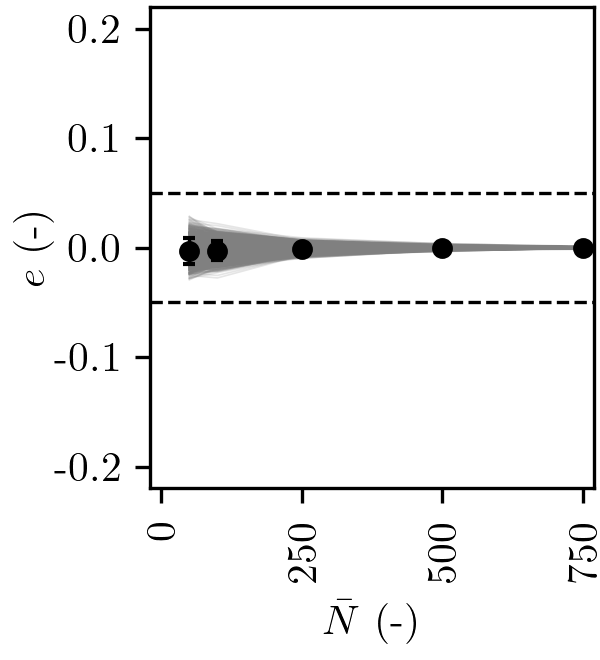}}
    \subfigure[]{%
        \label{subfig:strain_error_2}
         \includegraphics[width=2in]{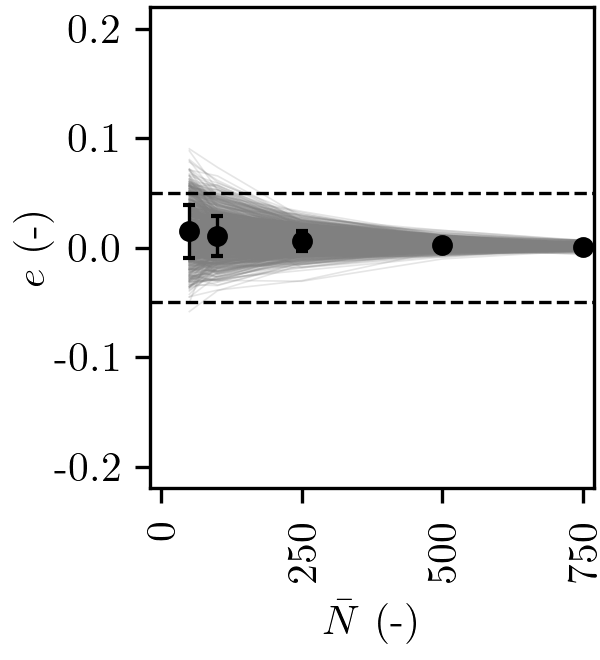}}
    \subfigure[]{%
        \label{subfig:strain_error_3}
        \includegraphics[width=2in]{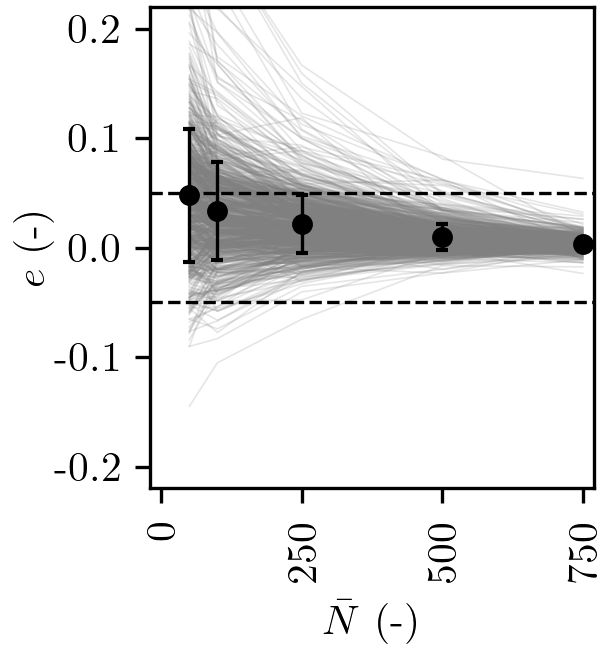}} 
    \caption{Grain-averaged equivalent elastic strain error metric, $e$, as a function of mesh density, $\bar{N}$, at macroscopic strains of \subref{subfig:strain_error_1} \SI{0.015}{\percent}, \subref{subfig:strain_error_2} \SI{0.2}{\percent}, and \subref{subfig:strain_error_3} \SI{10}{\percent}. Dashed lines bound the region within $\pm$\SI{5}{\percent} of the behavior predicted in the simulation performed with mesh density of $\bar{N}=1000$. Dots represent the average value at each mesh density, with whiskers representing $\pm \sigma_e$, where $\sigma_e$ is the standard deviation.}
    \label{fig:strain_error}
\end{figure}

\begin{figure}[htb!]
    \centering
    \subfigure[]{%
        \label{subfig:strain_error_metrics_1}
        \includegraphics[width=2in]{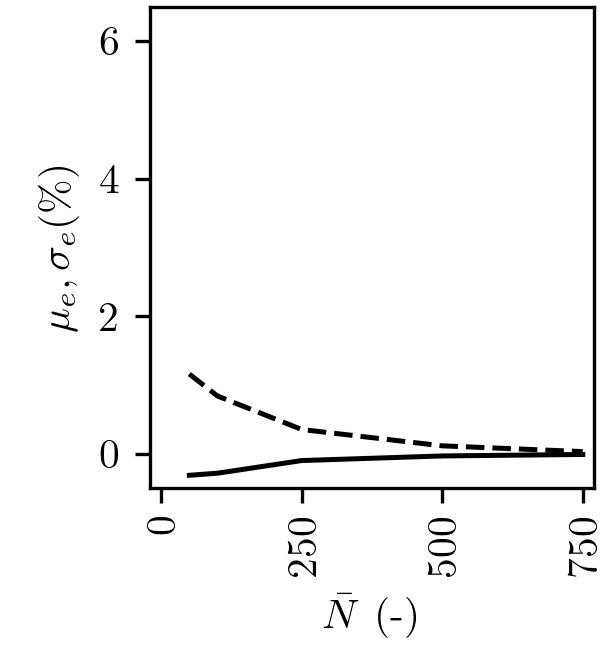}}
    \subfigure[]{%
        \label{subfig:strain_error_metrics_2}
         \includegraphics[width=2in]{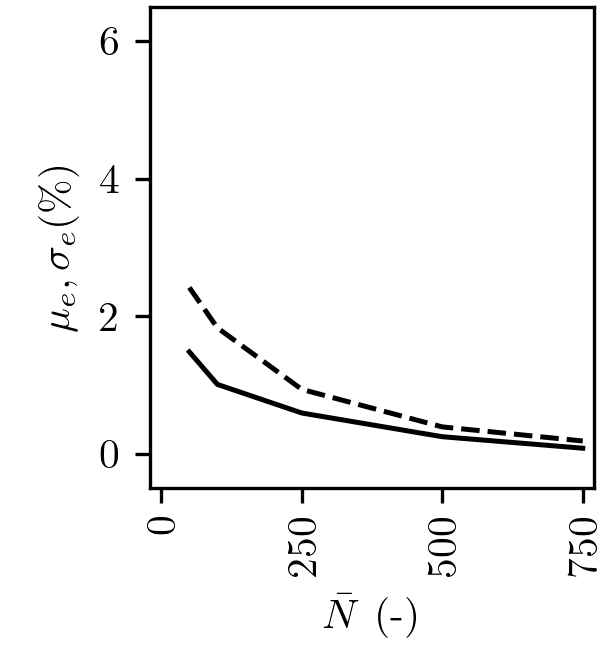}}
    \subfigure[]{%
        \label{subfig:strain_error_metrics_3}
        \includegraphics[width=2in]{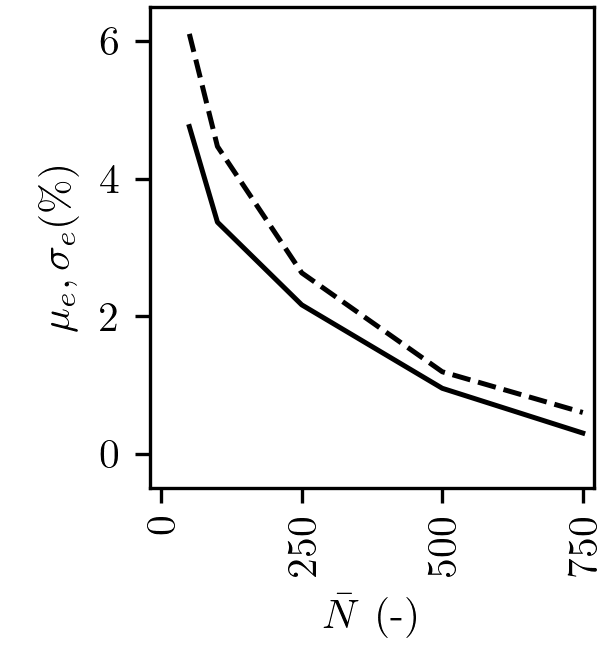}} 
    \caption{Mean ($\mu_e$, solid line) and standard deviation ($\sigma_e$, dashed line) of the grain-averaged equivalent elastic strain error metric, $e$, as a function of mesh density, $\bar{N}$, at macroscopic strains of \subref{subfig:strain_error_metrics_1} \SI{0.015}{\percent}, \subref{subfig:strain_error_metrics_2} \SI{0.2}{\percent}, and \subref{subfig:strain_error_metrics_3} \SI{10}{\percent}.}
    \label{fig:strain_error_metrics}
\end{figure}

For the $i$-th grain, we calculate a grain-averaged elastic strain tensor:
\begin{equation}
    \bar{\boldsymbol\varepsilon}_i^\text{e} = \sum_j^{M}{\frac{V^\text{elt}_j}{V^\text{grain}_i}\boldsymbol\varepsilon_j^\text{e}} \quad ,
    \label{eq:strain_el_vol_avg}
\end{equation}
where $M$ is the number of elements within the grain, $V^\text{elt}_j$ is the elemental volume and $\boldsymbol\varepsilon_j^\text{e}$ is the elemental elastic strain tensor of the $j$-th element within the grain, and $V^\text{grain}_i$ is the volume of the grain. From the grain-averaged elastic strain tensor, we calculate a grain-averaged equivalent elastic strain:
\begin{equation}
    \bar{\varepsilon}_i^\text{e} = \sqrt{\frac{2}{3}\bar{\boldsymbol\varepsilon}_i^\text{e}:\bar{\boldsymbol\varepsilon}_i^\text{e}} \quad .
    \label{eq:strain_el_eq}
\end{equation}
Finally, we define the grain-averaged elastic strain error for the $i$-th grain, $e_i$, a function of mesh density, $\bar{N}$, as:
\begin{equation}
    e_i\left(\bar{N}\right) = \frac{\bar{\varepsilon}_i^\text{e}(\bar{N}) - \hat{\varepsilon}_i^\text{e}}{\hat{\varepsilon}_i^\text{e}} \quad ,
    \label{eq:strain_el_eq_grain}
\end{equation}
where $\bar{\varepsilon}_i^\text{e}\left(\bar{N}\right)$ is the grain-averaged elastic strain for the $i$-th grain in the simulation with mesh density $\bar{N}$, and $\hat{\varepsilon}_i^\text{e}$ is the asymptotic value (i.e., the counterpart in the simulation with mesh density $\bar{N}=1000$). We plot the grain-averaged equivalent elastic strain error metric ($e_i$) for all grains in each of the 6 different meshes and at each of the macroscopic strain states of interest, and present these data in Figure~\ref{fig:strain_error}. We further present the isolated statistics---namely the averages and standard deviations of the grain-averaged equivalent elastic strain error across all grains as a function of mesh density---in Figure~\ref{fig:strain_error_metrics}.

\begin{figure}[htb!]
    \centering
    \subfigure[]{%
        \label{subfig:orientation_error_1}
         \includegraphics[width=2in]{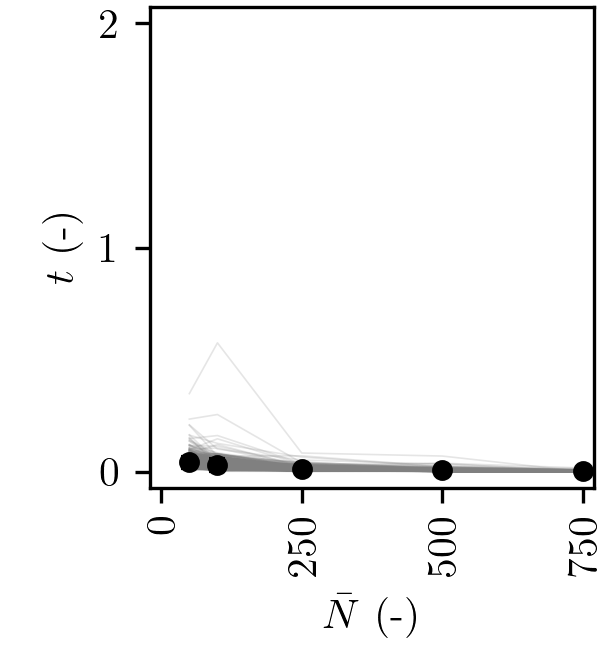}}
    \subfigure[]{%
        \label{subfig:orientation_error_2}
         \includegraphics[width=2in]{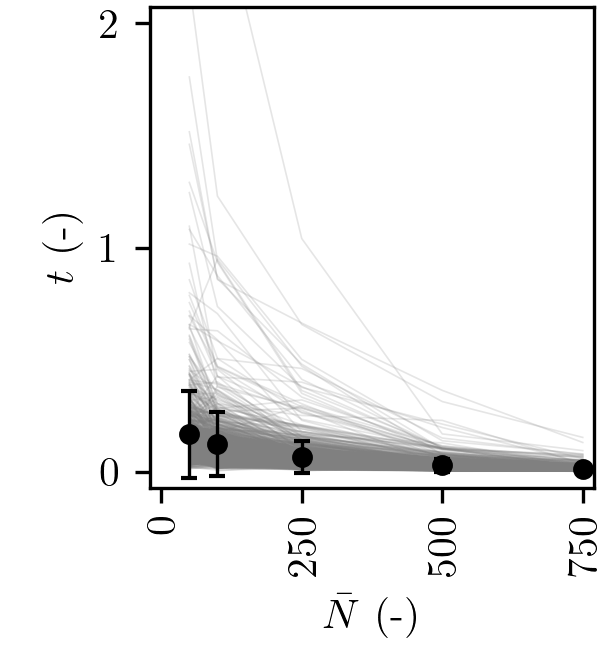}}
    \subfigure[]{%
        \label{subfig:orientation_error_3}
         \includegraphics[width=2in]{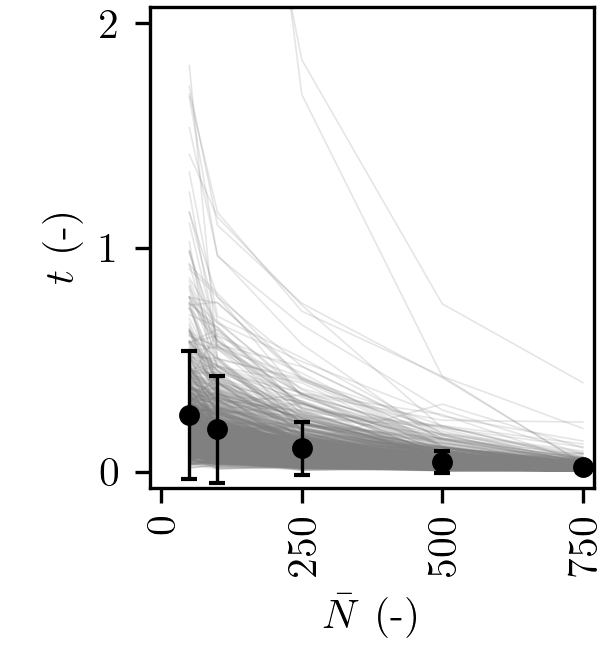}}         
    \caption{Grain-averaged orientation error metric, $t$, as a function of mesh density, $\bar{N}$, at macroscopic strains of \subref{subfig:orientation_error_1}~\SI{0.015}{\percent}, \subref{subfig:orientation_error_2} \SI{0.2}{\percent}, and \subref{subfig:orientation_error_3} \SI{10}{\percent}. Dots represent the average value at each mesh density, with whiskers representing $\pm \sigma_t$, where $\sigma_t$ is the standard deviation.}
    \label{fig:orientation_error}
\end{figure}

\begin{figure}[htb!]
    \centering
    \subfigure[]{%
        \label{subfig:orientation_error_metrics_1}
        \includegraphics[width=2in]{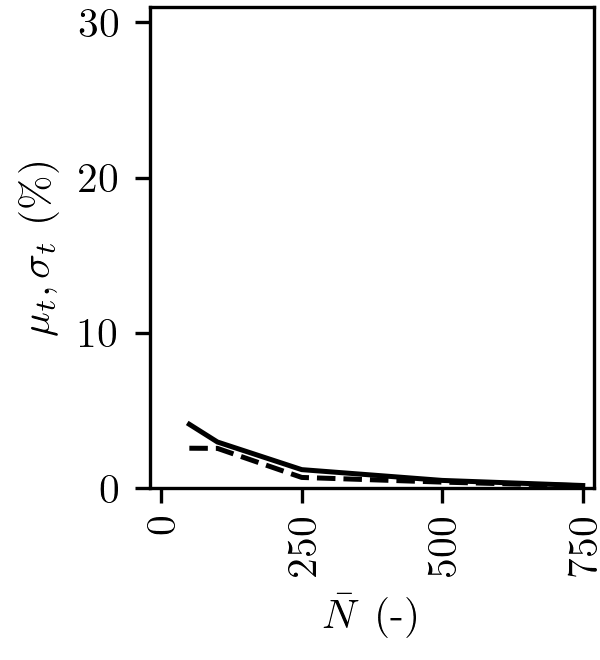}}
    \subfigure[]{%
        \label{subfig:orientation_error_metrics_2}
         \includegraphics[width=2in]{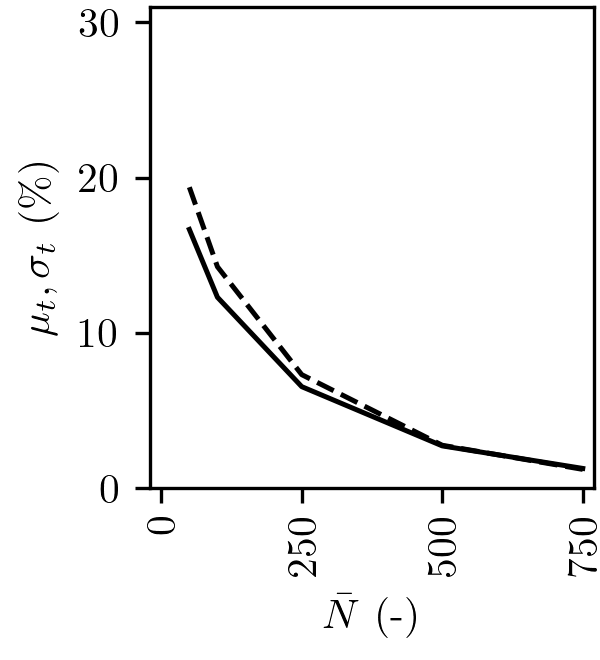}}
    \subfigure[]{%
        \label{subfig:orientation_error_metrics_3}
        \includegraphics[width=2in]{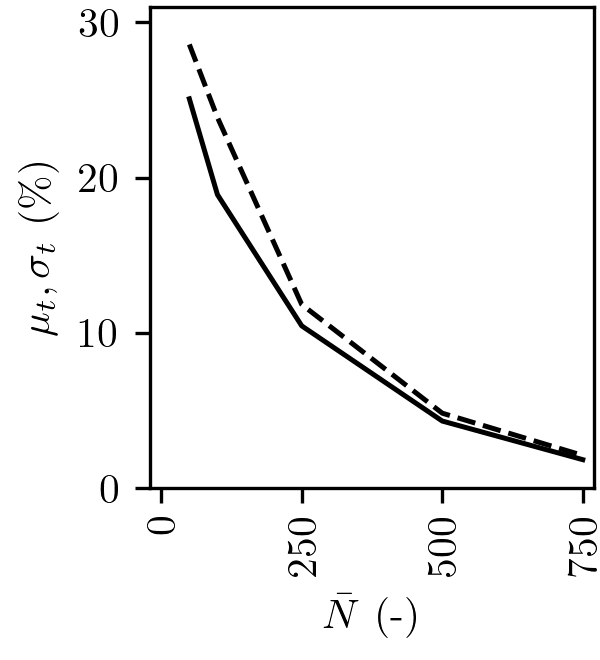}} 
    \caption{Mean ($\mu_t$, solid line) and standard deviation ($\sigma_t$, dashed line) of the grain-averaged orientation error metric, $t$, as a function of mesh density, $\bar{N}$, at macroscopic strains of \subref{subfig:orientation_error_metrics_1} \SI{0.015}{\percent}, \subref{subfig:orientation_error_metrics_2} \SI{0.2}{\percent}, and \subref{subfig:orientation_error_metrics_3} \SI{10}{\percent}.}
    \label{fig:orientation_error_metrics}
\end{figure}

Finally, we inspect the behavior of the other primary metric reconstructed from HEDM data: the grain-averaged orientation. To begin, we compute the grain-averaged orientation, $\bar{\mathbf{q}}$, from the elemental values of the grain, by quaternion averaging~\cite{glez}. Next, we generally define the misorientation angle, $\theta\left(\mathbf{q}_1,\mathbf{q}_2\right)$, between two orientations, $\mathbf{q}_1$ and $\mathbf{q}_2$, as the angle associated with the minimum rotation that brings one orientation into coincidence with the other, and we compute as:
\begin{equation}
    \theta\left(\mathbf{q}_1,\mathbf{q}_2\right) = \underset{i}{\mathrm{min}} \left( 2 \, \arccos \left| \left(\mathbf{q}_2 \, \mathbf{u}_i \, \mathbf{q}_1^{-1} \right)_0 \right| \right) \quad ,
\end{equation}
where `$\left(...\right)_0$' is the scalar part of a quaternion, and $\left\{\mathbf{u_i},\,i \in \left[1, n\right]\right\}$ are the operators of the crystal symmetry group. Finally, we define a grain-averaged orientation error metric, $t$, a function of mesh density, $\bar{N}$, as:
\begin{equation}
    t_i\left(\bar{N}\right) = \frac{\theta_i\,(\bar{\mathbf{q}}_i (\bar{N}),\,\hat{\mathbf{q}}_i)}{\theta_i\left(\hat{\mathbf{q}}_i^0,\,\hat{\mathbf{q}}_i\right)} \quad,
\end{equation}
where $\theta_i\,(\bar{\mathbf{q}}_i (\bar{N}),\,\hat{\mathbf{q}}_i)$ is the misorientation between the grain-averaged orientations of the $i$-th grain in the simulation with mesh density $\bar{N}$ and the asymptotic value (i.e., its counterpart in the simulation with mesh density $\bar{N}=1000$), and $\theta_i(\hat{\mathbf{q}}_i^0,\,\hat{\mathbf{q}}_i)$ is the asymptotic grain-averaged reorientation of the $i$-th grain (i.e., the grain-averaged reorientation of the grain in the simulation with mesh density $\bar{N}=1000$ compared to its initial orientation, $\hat{\mathbf{q}}^0$). We plot the orientation error metric for all grains in each of the six different meshes and across multiple macroscopic strain states in Figure~\ref{fig:orientation_error}, as well as the isolated statistics in Figure~\ref{fig:orientation_error_metrics}.

\section{Discussion}
\label{sec:discussion}

\subsection{Macroscopic Behavior}
\label{sec:macro_behavior}

We first provide comment on the results presented in Section~\ref{sec:results}. Inspecting the macroscopic behavior presented in Figure~\ref{fig:stress_strain}, we observe that the stress decreases as the mesh density increases. We calculate macroscopic stresses of approximately \SI{23.7}{\mega\pascal}, \SI{53.8}{\mega\pascal}, and \SI{218.6}{\mega\pascal} at \SI{0.015}{\percent}, \SI{0.2}{\percent}, and \SI{10}{\percent} macroscopic strain, respectively, for the simulation performed with a mesh density of $\bar{N}=50$, and \SI{23.4}{\mega\pascal}, \SI{51.7}{\mega\pascal}, and \SI{205.2}{\mega\pascal} at the same macroscopic states for the simulation performed with a mesh density of $\bar{N}=1000$ (i.e., differences of \SI{0.3}{\mega\pascal}, \SI{2.1}{\mega\pascal}, and \SI{13.4}{\mega\pascal}, respectively). We further notice a diminishing decrease in stress as $\bar{N}$ is increased: i.e., there is a larger drop in stress when increasing the mesh density from $\bar{N}=50$ to $\bar{N}=100$ (e.g., \SI{3.8}{\mega\pascal} at \SI{10}{\percent} macroscopic strain) than compared to that exhibited when increasing the mesh density from $\bar{N}=750$ to $\bar{N}=1000$ (i.e., \SI{0.7}{\mega\pascal} at \SI{10}{\percent} macroscopic strain), indicating asymptotic or converged behavior in the macroscopic response. This is seen visually in Figure~\ref{subfig:stress_strain_2}, where the response of the $\bar{N}=750$ simulation nearly overlaps with that of the $\bar{N}=1000$ simulation.

\subsection{Grain-Averaged State}
\label{sec:grain_avg_state}
We next investigate the predictions of the grain-averaged equivalent elastic strain as a function of mesh density. Figure~\ref{fig:mesh_strain} depicts the grain-averaged equivalent elastic strain plotted spatially on the deformed finite element mesh at the three macroscopic states of interest. We note the fundamentally similar spatial distributions of elastic strains between the $\bar{N}=50$ and $\bar{N}=1000$ cases, which indicates that even the errors resulting from a coarse mesh ($\bar{N}=50$) remain significantly lower than the typical variations of the elastic strain from one grain to the other. For a more quantitative analysis, we can first note that the grain-to-grain variations of the grain-averaged elastic strains can be described by the coefficient of variation, i.e., the ratio between the standard deviation and the mean value. This ratio has values (computed for the reference $\bar{N}=1000$ case) of \SI{16}{\percent}, \SI{26}{\percent} and \SI{26}{\percent} at macroscopic strains of~\SI{0.015}{\percent}, \SI{0.2}{\percent}, and \SI{10}{\percent}, respectively. 

In parallel, we inspect the error in the grain-averaged equivalent elastic strain predictions, $e$, as a function of both macroscopic state and mesh density, as presented in Figures~\ref{fig:strain_error} and~\ref{fig:strain_error_metrics}. Here, we first see that simulations performed with coarser mesh densities tend to introduce a systematic bias (non-zero $\mu_e$ values), and further that deviations tend to favor higher strains in the simulations performed with coarser mesh densities. We find that both observations correlate with the behavior observed at the macroscopic scale: the former indicates convergent behavior as mesh density is increased, and the latter indicates higher stresses in simulations performed with coarser mesh densities. We also find generally higher error magnitudes as a function of a decrease in mesh density at all macroscopic states considered, though the error increases greatly as a function of the macroscopic state. It is, however, interesting to compare the error values to the coefficient of variations (between grains). For $\bar{N}=50$, the standard deviation of the error ($\sigma_e$, disregarding the systematic bias ($\mu_e$)) is equal to \SI{1.2}{\percent}, \SI{2.4}{\percent} and \SI{6.1}{\percent}, respectively, which remains smaller than the coefficients of variations by factors ranging from 4 to 14.  For $\bar{N}\geq250$,  the values of $\sigma_e$ fall below \SI{1}{\percent} at macroscopic strains of~\SI{0.015}{\percent} and \SI{0.2}{\percent}, and \SI{2.6}{\percent} at~\SI{10}{\percent} macroscopic strain. These values are 10 to 46 times smaller than the coefficients of variation.

We observe similar behavior in the prediction of the grain-averaged orientations. In particular, we find that simulations performed with coarser mesh density tend to exhibit higher error in the grain-averaged orientation predictions, $t$, as shown in Figure~\ref{fig:orientation_error}. Again, we observe that the error is low considering deformation in the elastic regime, and increases drastically during the evolution of macroscopic plasticity. Likewise, we provide statistics regarding the mean error and the spread in error in Figure~\ref{fig:orientation_error_metrics}. We again observe convergent behavior.

Collectively, we find that error in both the prediction of the grain-averaged equivalent plastic strain and in the grain-averaged orientation---and in particular the mean and standard deviation of the error distributions---is generally low in the elastic regime, sharply increases during the onset of plasticity, and compounds as plasticity continues to develop. We attribute this to the path-dependent nature of the plastic deformation response of polycrystalline materials, and note that small deviations early in the deformation history may compound and lead to large errors later in the deformation history. However, and particularly for the grain-averaged elastic strains, the error remain significantly lower than the coefficient of variation (among grains), by factors of about 5 for $\bar{N}=50$ and higher than 10 for $\bar{N}\geq250$.

It is worth placing the presented results in context of the experimental data with which they will be compared. HEDM results, and nearly all elastic strain or stress measurements, are reported in terms of absolute values \cite{noyandef}. In ideal cases, the per component uncertainties for elastic strains measured using ff-HEDM can be less than $10^{-4}$~\cite{hurley2018characterization} and misorientation of \SI{0.01}{\degree}~\cite{poulsen2004three}. However, these best case scenarios are usually associated with well-annealed grains with minimal dislocation content. As grains deform and diffraction peaks broaden, uncertainties can approach $3 \times 10^{-4}$. In general, as applied stresses (and corresponding elastic strains) increase, we note the likelihood for errors in the simulation to exceed the uncertainties in the experiment will grow.

\subsection{Dependence on Crystallographic Orientation}
\label{sec:ori_effect}

\begin{figure}[htb!]
    \centering
     \subfigure[]{%
        \label{subfig:ipf_errors_1}
        \includegraphics[height=2in]{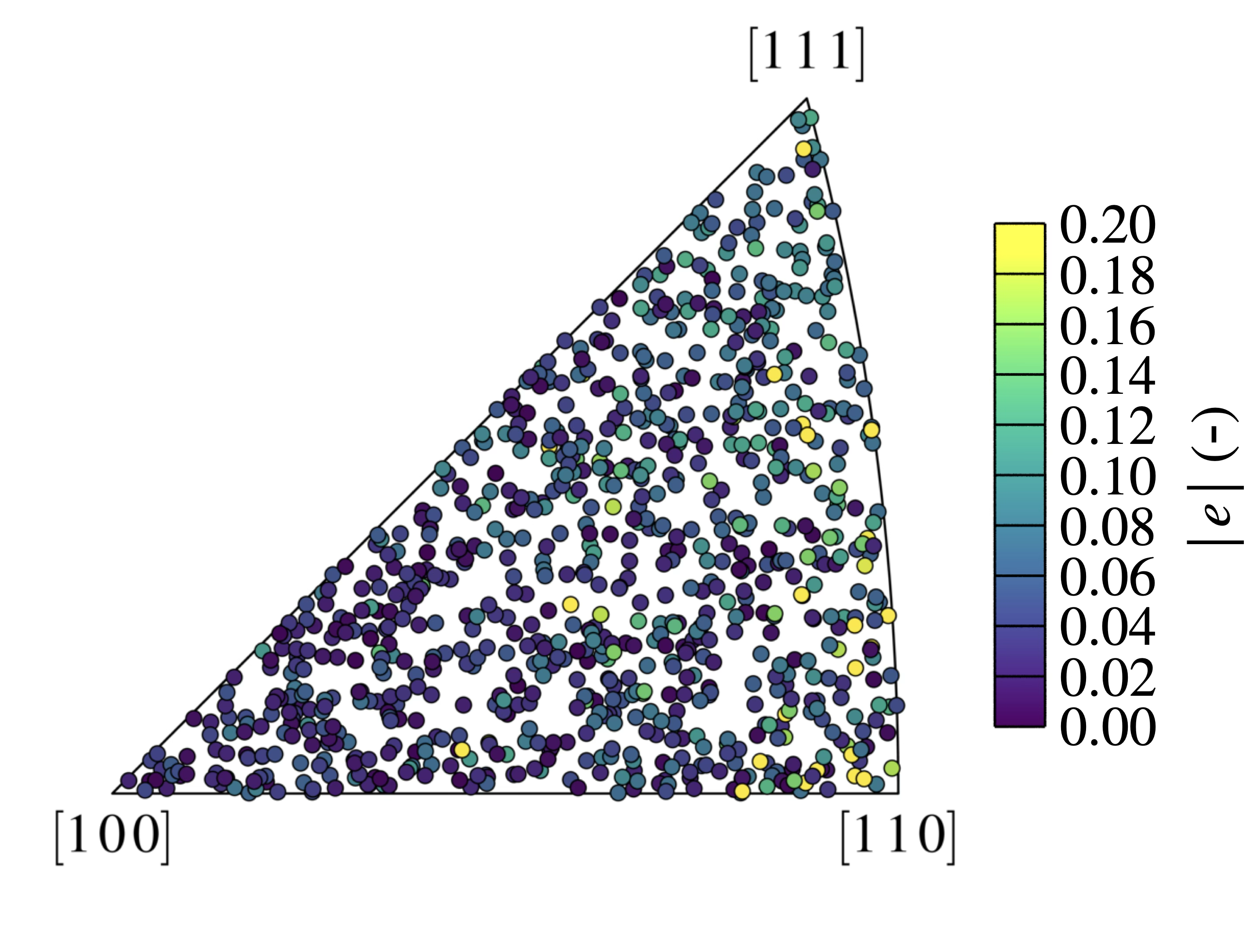}} 
    \subfigure[]{%
        \label{subfig:ipf_errors_2}
        \includegraphics[height=2in]{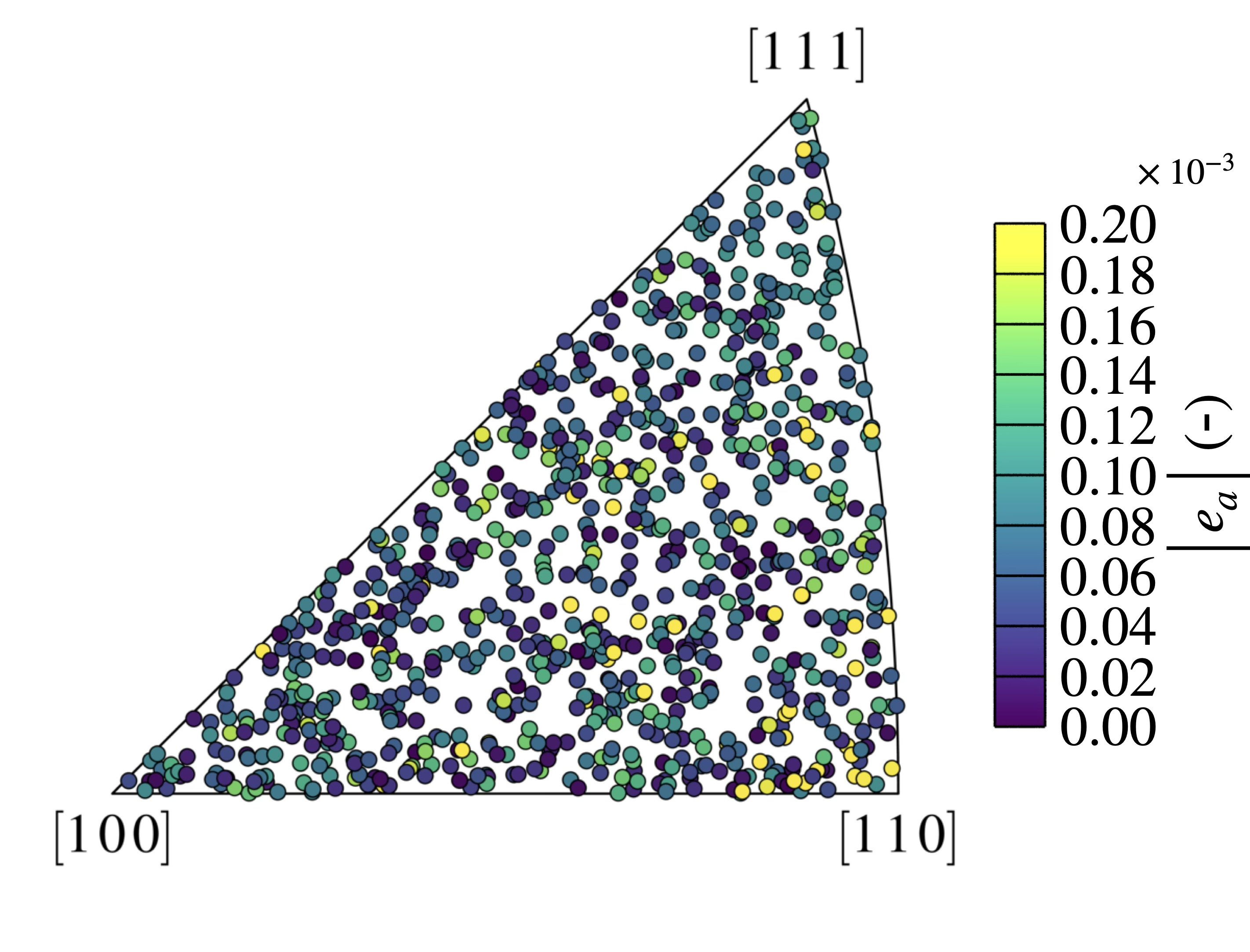}}
    \caption{Inverse pole figures with respect to the sample $z$ direction, depicting~\subref{subfig:ipf_errors_1} the magnitude of the grain-averaged equivalent elastic strain error metric, $e$, and~\subref{subfig:ipf_errors_2} the magnitude of the absolute error (i.e., the magnitude of the numerator of Equation~\ref{eq:strain_el_eq_grain}), both plotted considering the initial orientations of the grains.}
    \label{fig:ipf_errors}
\end{figure}

We next analyze the dependency of the error on the initial orientation of the grain. Particularly, we consider the absolute value of the error metric, $|e|$, and plot as a function of the initial orientations of the grains on an inverse pole figure, as depicted in Figure~\ref{subfig:ipf_errors_1} (we choose here, for example, the errors between the $\bar{N}=50$ and $\bar{N}=1000$ cases at \SI{10}{\percent} macroscopic strain).
We note a relatively homogeneous distribution of errors as a function of orientation, indicating a relatively weak effect due to crystallographic orientation. However, we do note that orientations which lie along the crystallographic fiber composed of crystals with their \hkl<0 0 1> direction aligned with the axis of extension / the sample $z$ direction (or the \hkl<0 0 1>$||z$ fiber) display somewhat lower errors than other orientations, and a potential bias towards high error along the \hkl<1 1 0>$||z$ and \hkl<1 1 1>$||z$ fibers. Our results echo those found in ref.~\cite{Lim2019}, which display similar trends regarding the average of the elemental-scale errors. Figure~\ref{subfig:ipf_errors_2} represents the distribution of the magnitude of the absolute error, $\left|e_a\right|$, which corresponds to the magnitude of the numerator of Equation~\ref{eq:strain_el_eq_grain}. Interestingly, the absolute error shows very weak correlation to orientation, with low and high values equally distributed over the standard triangle. Mechanically, we therefore attribute the slight biases in absolute relative error ($e$) to the slip behavior of the grains along these fibers: those along the \hkl<1 1 0>$||z$ and \hkl<1 1 1>$||z$ fibers are more favorably oriented for crystallographic slip (i.e., ``soft'' grains), while those oriented along the \hkl<0 0 1>$||z$ fiber are comparably not (i.e., ``hard'' grains). We note, however, that this bias is dominated by large variations of $e$ even between grains of similar orientations, especially for orientations away from the \hkl<0 0 1>$||z$ fiber.

\subsection{Worst-Case Grain and Traction-Free Surfaces}
\label{sec:worst_grain}

\begin{figure}[htb!]
    \centering
    \subfigure[]{%
        \label{subfig:worst_grain_spatial_1}
        \includegraphics[height=1.6in]{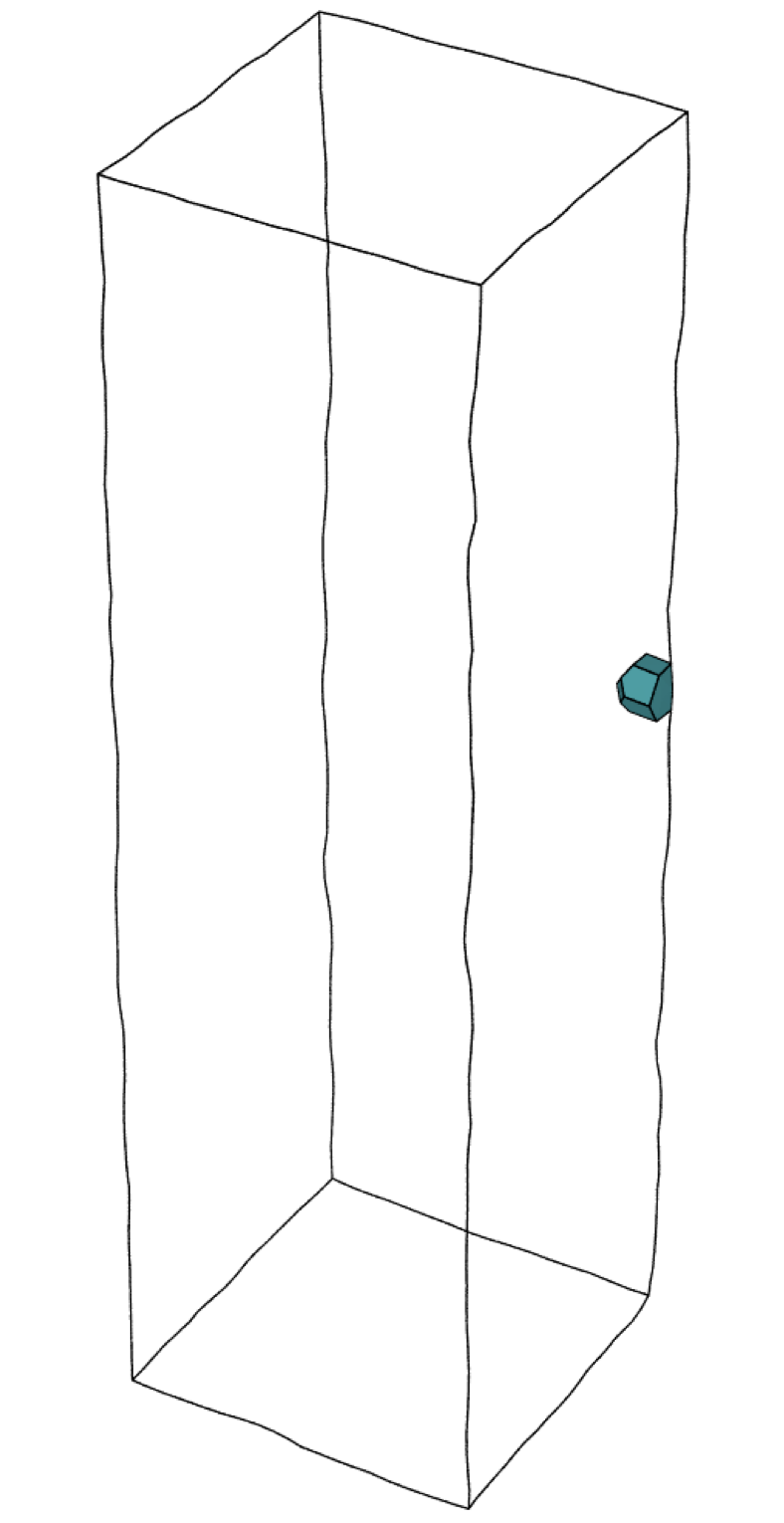}} 
    \subfigure[]{%
        \label{subfig:worst_grain_spatial_2}
        \includegraphics[height=1.6in]{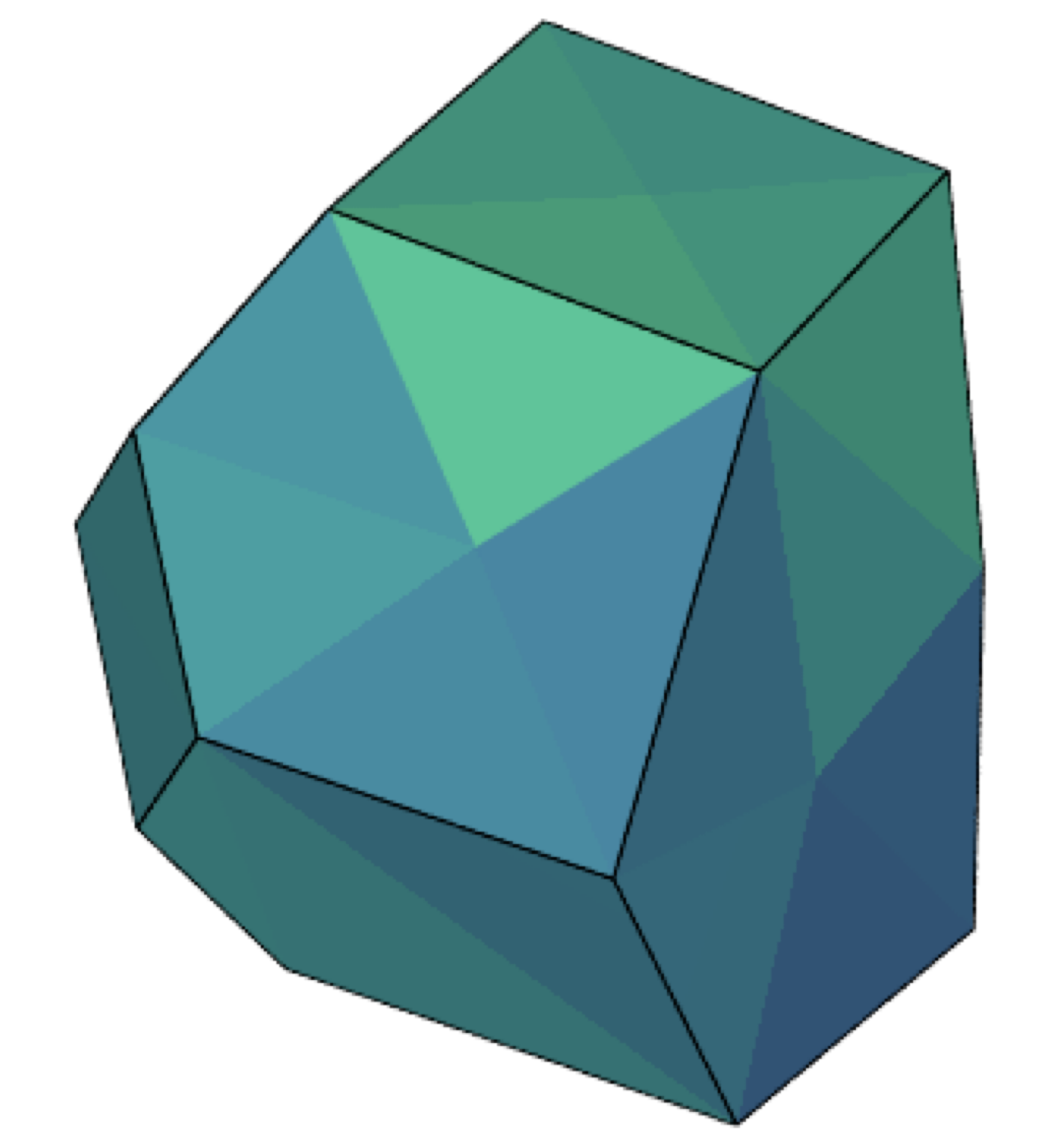}}
    \subfigure[]{%
        \label{subfig:worst_grain_spatial_3}
        \includegraphics[height=1.6in]{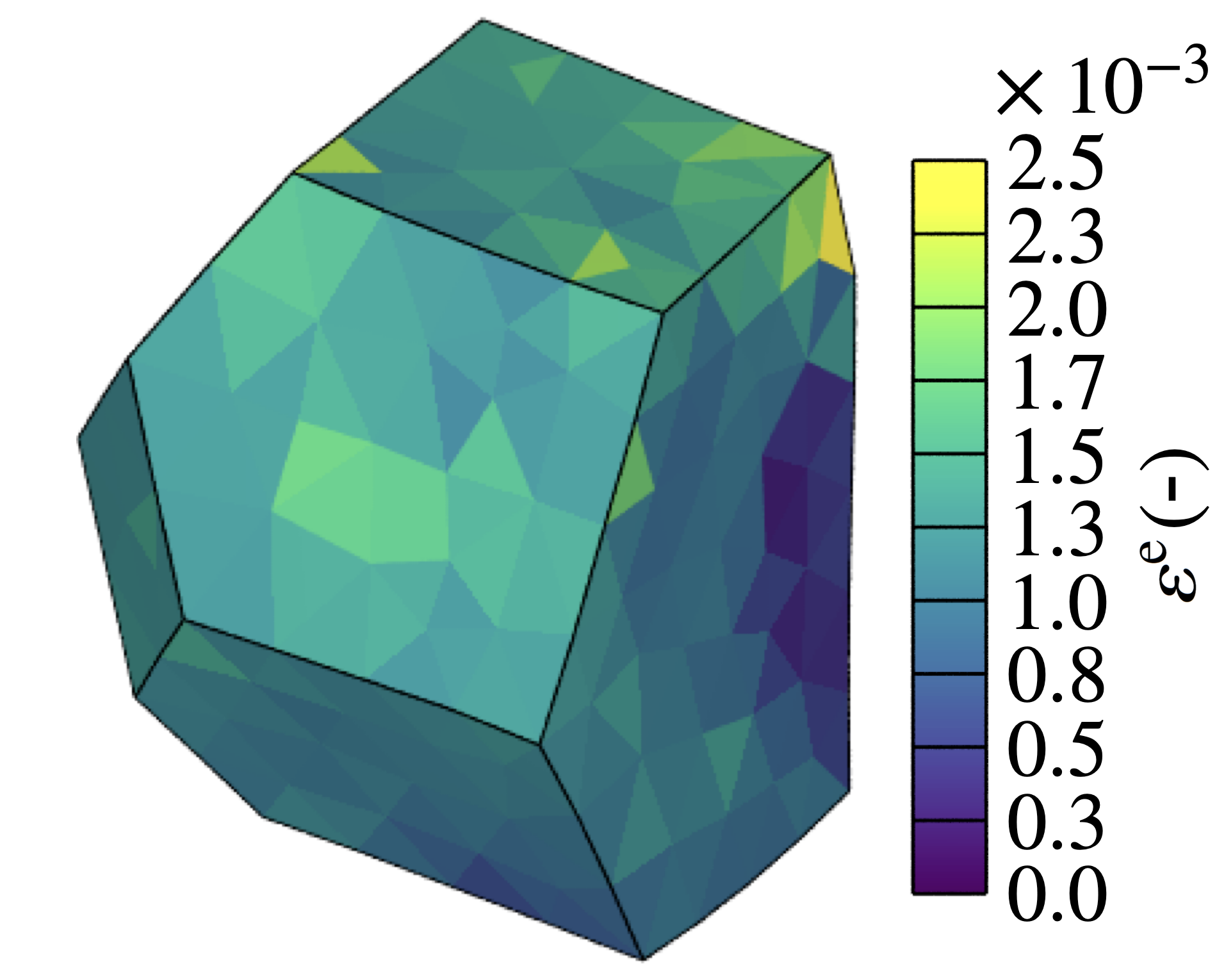}}
    \caption{Visualizations of the grain with the greatest value of $e$ at \SI{10}{\percent} macroscopic strain, where~\subref{subfig:worst_grain_spatial_1} depicts the grain's spatial location in the polycrystal, while~\subref{subfig:worst_grain_spatial_2} and~\subref{subfig:worst_grain_spatial_3} depict the elemental-scale equivalent elastic strain, $\varepsilon^\text{e}$, plotted spatially on the deformed meshes for the simulations performed with mesh densities of $\bar{N}=50$ and $\bar{N}=1000$, respectively.}
    \label{fig:worst_grain_spatial}
\end{figure}

\begin{figure}[htb!]
    \centering
    \subfigure[]{%
        \label{subfig:worst_grain_statistical_1}
         \includegraphics[width=2in]{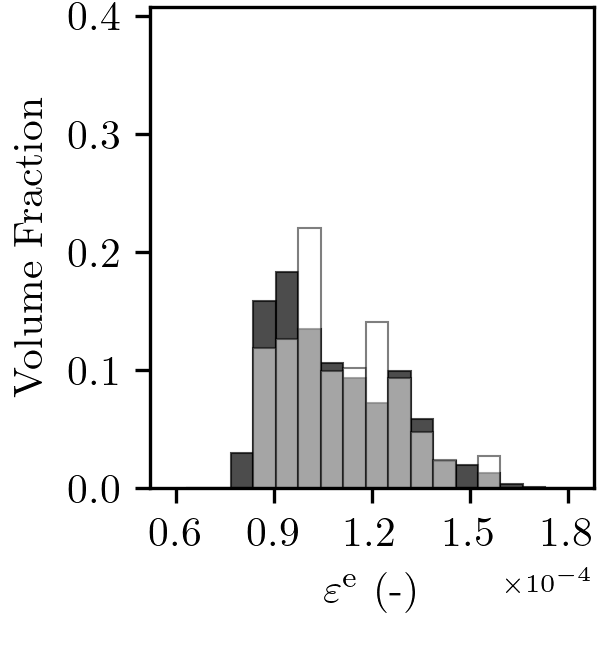}}
    \subfigure[]{%
        \label{subfig:worst_grain_statistical_2}
         \includegraphics[width=2in]{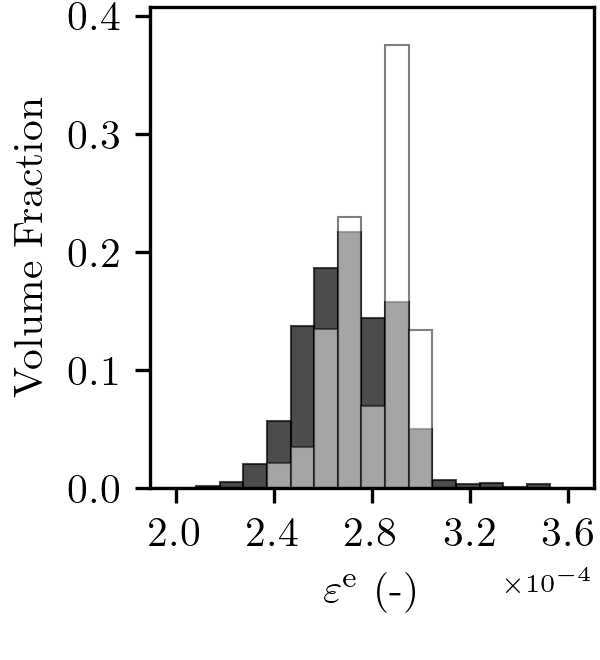}} 
    \subfigure[]{%
        \label{subfig:worst_grain_statistical_3}
         \includegraphics[width=2in]{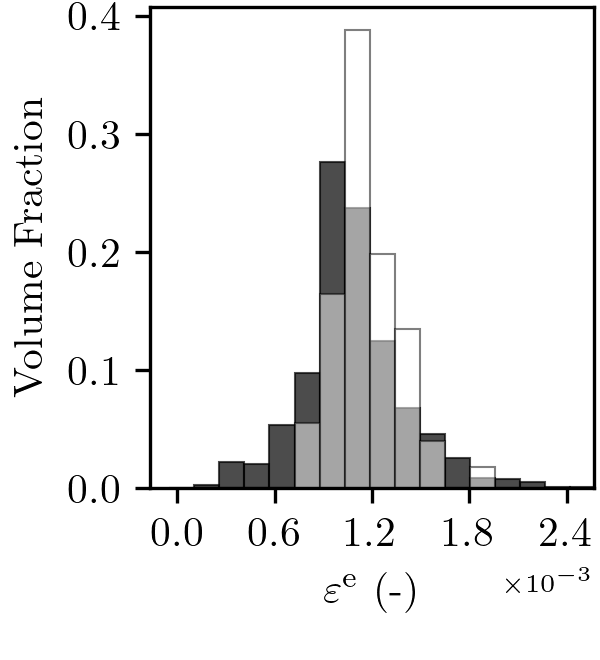}}
    \caption{Statistical distributions of the elemental-scale  elastic strain, $\varepsilon^\text{e}$, for the grain with the greatest values of $e$, with data from simulations performed with mesh densities of $\bar{N}=50$ (white bars) and $\bar{N}=1000$ (dark gray bars) at macroscopic strains of~\subref{subfig:worst_grain_statistical_1} \SI{0.015}{\percent},~\subref{subfig:worst_grain_statistical_2} \SI{0.2}{\percent}, and~\subref{subfig:worst_grain_statistical_3} \SI{10}{\percent}. Note the difference in strain scale between the three figures.}
    \label{fig:worst_grain_statistical}
\end{figure}

We next inspect the grain which exhibits the greatest magnitude of strain error (i.e., the limiting case) in order to highlight the underlying trends influencing the differences in average state. To this end, we plot the elemental scale equivalent elastic strain, $\varepsilon^\text{e}$, spatially on the deformed grain at \SI{10}{\percent} macroscopic strain for the two limiting mesh densities in Figure~\ref{fig:worst_grain_spatial}. We further present histograms displaying the statistical distribution (volume fractions) of strain across the grains in Figure~\ref{fig:worst_grain_statistical}.

Qualitatively, we note that the grain with the coarser mesh refinement fails to capture the more complex gradient in the equivalent elastic strain field exhibited by the grain from asymptotic case. While the grain from the simulation with density $\bar{N}=50$ shows broadly similar behavior, a large region of low elastic strain is entirely absent. This behavior is confirmed via quantitative inspection of the statistical distributions of equivalent elastic strain, where we find both that the grain from the simulation with mesh density $\bar{N}=50$ exhibits generally higher strains than that from the simulation with mesh density $\bar{N}=1000$, and especially a more pronounced distribution tail towards lower strains in the latter case. We attribute this overall behavior to the lack of mesh fidelity in regions of significant gradients in the deformation field. In this particular example, we note that the portion of the grain in the $\bar{N}=50$ case that fails to capture the low elastic strain region of the field occurs at a vertex point of two elements, and that there are only two elements which span the edge of the grain. In other words, there are limited degrees of freedom which consequently disallow for the capture of the actual deformation field.

\begin{figure}[htb!]
    \centering
    \subfigure[]{%
        \label{subfig:compare_strain_error_metrics_1}
        \includegraphics[width=2in]{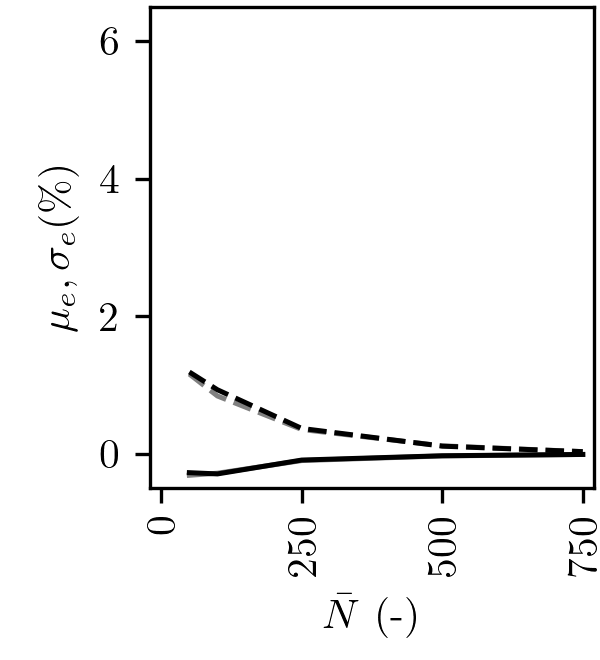}}
    \subfigure[]{%
        \label{subfig:compare_strain_error_metrics_2}
         \includegraphics[width=2in]{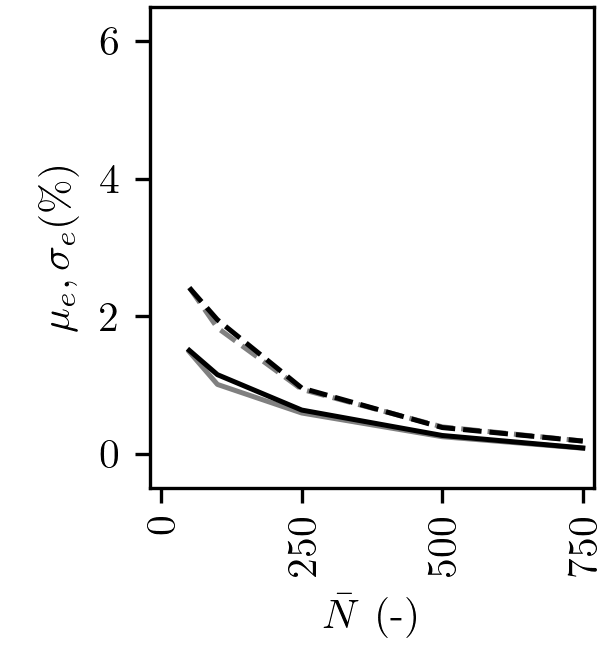}}
    \subfigure[]{%
        \label{subfig:compare_strain_error_metrics_3}
        \includegraphics[width=2in]{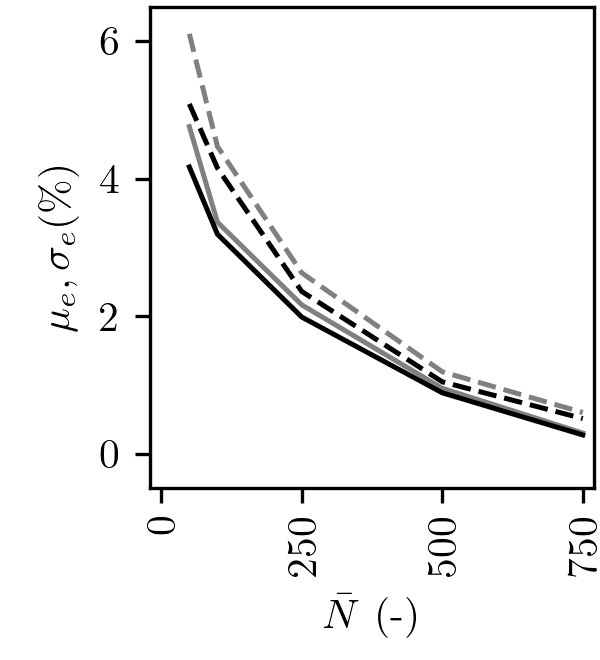}} 
    \caption{Mean ($\mu_e$, solid line) and standard deviation ($\sigma_e$, dashed line) of the grain-averaged equivalent elastic strain error metric, $e$, as a function of mesh density, $\bar{N}$, at macroscopic strains of \subref{subfig:compare_strain_error_metrics_1} \SI{0.015}{\percent}, \subref{subfig:compare_strain_error_metrics_2} \SI{0.2}{\percent}, and \subref{subfig:compare_strain_error_metrics_3} \SI{10}{\percent}. Black lines represent the statistics of the subset of grains in the interior of the sample, while gray lines represent the statistics from all grains in the central region (i.e., a reproduction of the data in Figure~\ref{fig:strain_error_metrics}).}
    \label{fig:compare_strain_error_metrics}
\end{figure}

We find it interesting to note that the location of the worst-case grain is on a traction-free surface of the domain. We thus question whether the traction-free surfaces lead to any influence on the observed behavior: i.e., do the grains which appear on the traction-free surface of the domain exhibit higher errors overall, or is the appearance of the worst-case grain on the traction-surface merely a coincidence? To address this, we perform the same analysis regarding the isolation of the statistics of the grain-averaged equivalent elastic strain error metric, $e$, as we presented for the entire set of grains in Figure~\ref{fig:strain_error_metrics} above, but here for only the interior grains (which number 676, compared to the 912 grains in the entire set above). We present this data in Figure~\ref{fig:compare_strain_error_metrics}.

Overall, we find that the overall behavior (in terms of error) on the interior of the domain is not appreciably different than considering all grains within the set considered. Indeed, at relatively low macroscopic strains, we find that the data nearly perfectly overlaps, indicating that there is almost no effect on the overall behavior. In contrast, we do observe some difference at higher macroscopic strain, namely a low---though apparent---drop in both the average and standard deviation of the error when considering only the interior grains. This is most evident at the lowest mesh density, and becomes almost indistinguishable as mesh densities increase. Collectively, we thus find that at most macroscopic states and mesh densities, the surface grains do not appear to have higher than average errors, as the overall statistics are not appreciably changed when they are removed from analysis.

\subsection{Effects of Boundary Conditions}
\label{sec:bcs}

\begin{figure}[htb!]
    \centering
    \subfigure[]{%
        \label{subfig:mesh_50_bcs_newbcs}
        \includegraphics[width=2.5in]{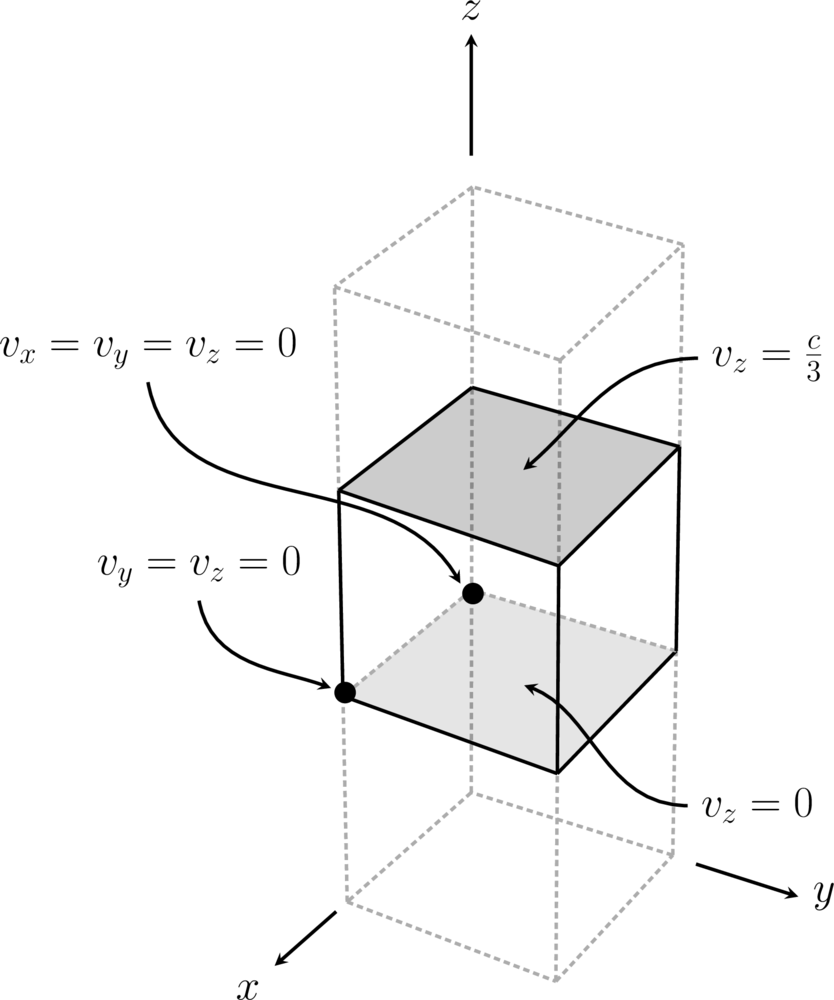}}
    \subfigure[]{%
        \label{subfig:mesh_50_bcs}
        \raisebox{0.5in}{\includegraphics[width=0.29\textwidth]{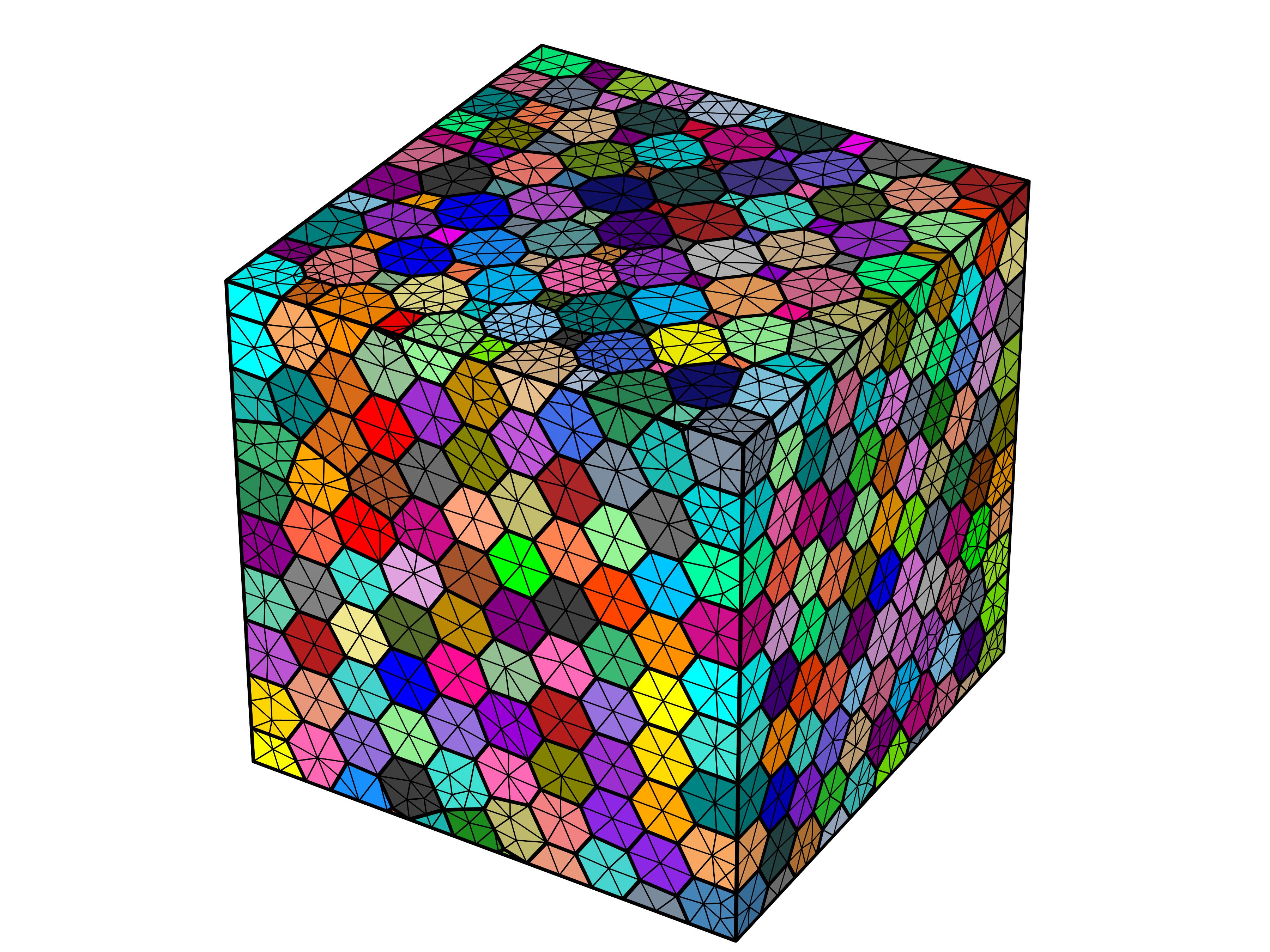}}}
    \subfigure[]{%
        \label{subfig:mesh_50_bcs_reduced}
        \raisebox{0.5in}{\includegraphics[width=0.29\textwidth]{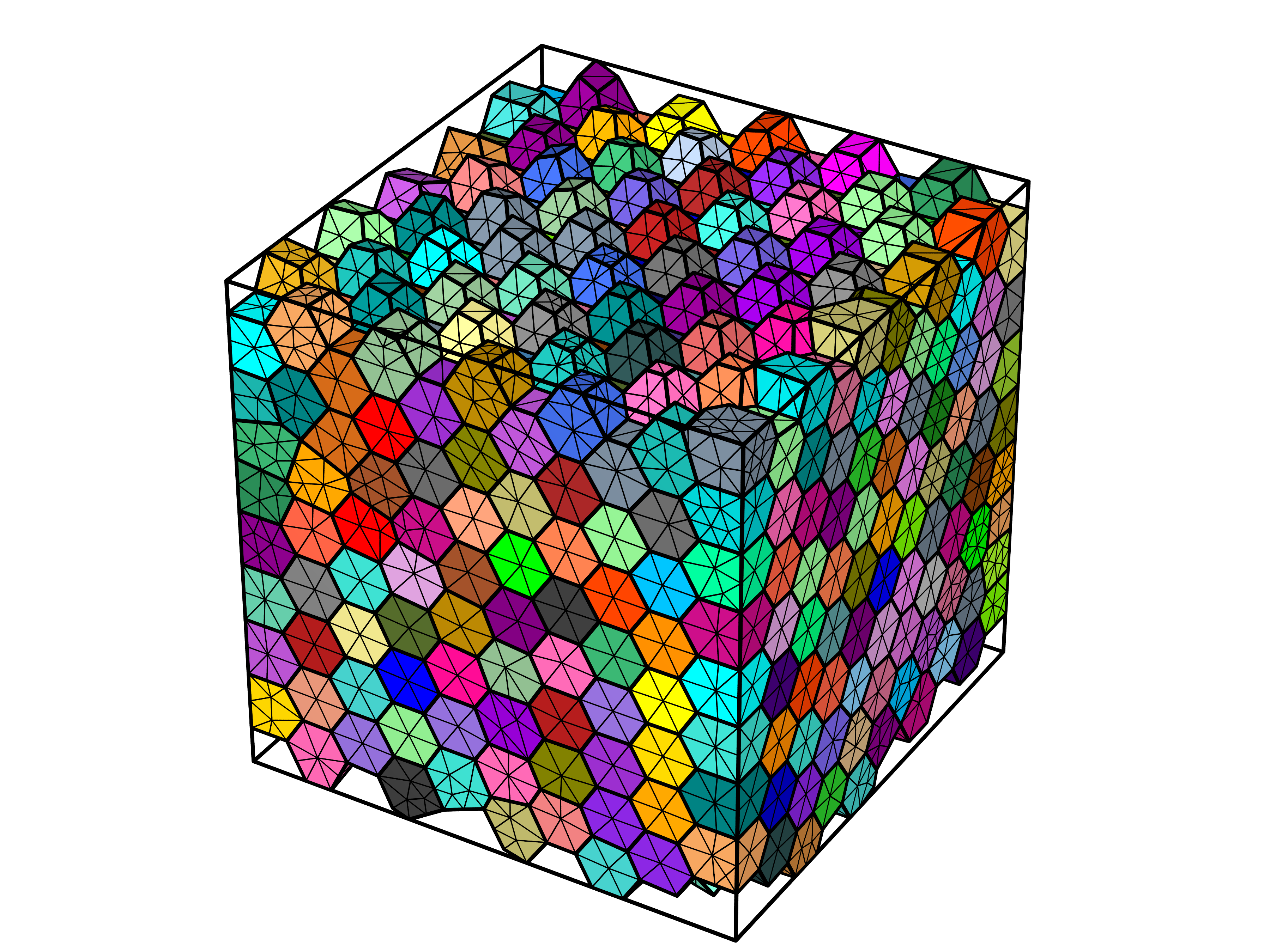}}}
    \caption{Visualizations of the boundary conditions and tessellation representing the polycrystal utilized in the boundary conditions analysis. \subref{subfig:mesh_50_bcs_newbcs} Boundary conditions utilized in the simulations (compare to Figure~\ref{fig:minimal_bc}), further detailing the location of the domain with reduced aspect ratio within the initial domain. \subref{subfig:mesh_50_bcs}~Tessellation used for the simulation, and~\subref{subfig:mesh_50_bcs_reduced} interior grains, considered for the analysis. The grains in the polycrystal are colored arbitrarily. Figures~\subref{subfig:mesh_50_bcs} and \subref{subfig:mesh_50_bcs_reduced} are provided for $\bar{N}=50$.}
    \label{fig:mesh_bcs}
\end{figure}

\begin{figure}[htb!]
    \centering
    \subfigure[]{%
        \label{subfig:strain_error_bcs_domain_compare_scatter_1000_1}
        \includegraphics[width=2in]{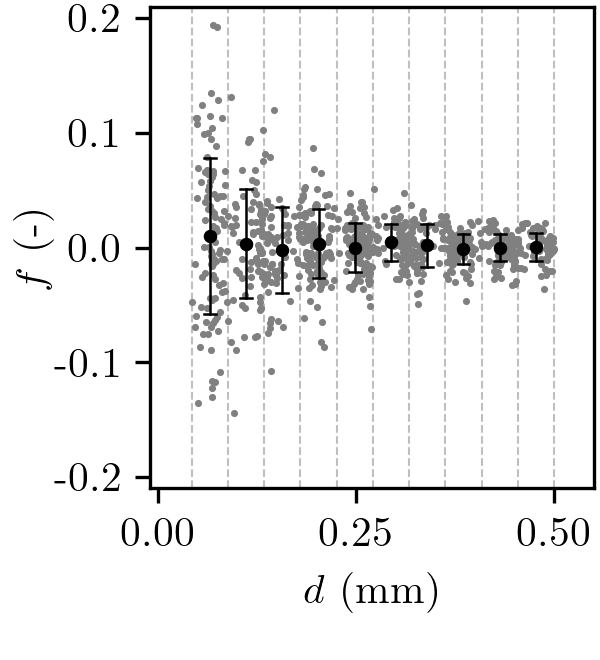}}
    \subfigure[]{%
        \label{subfig:strain_error_bcs_domain_compare_scatter_1000_2}
         \includegraphics[width=2in]{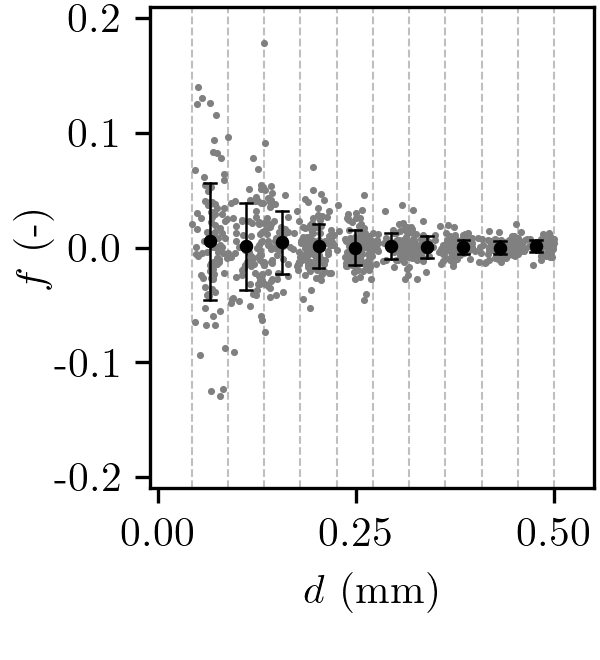}}
    \subfigure[]{%
        \label{subfig:strain_error_bcs_domain_compare_scatter_1000_3}
        \includegraphics[width=2in]{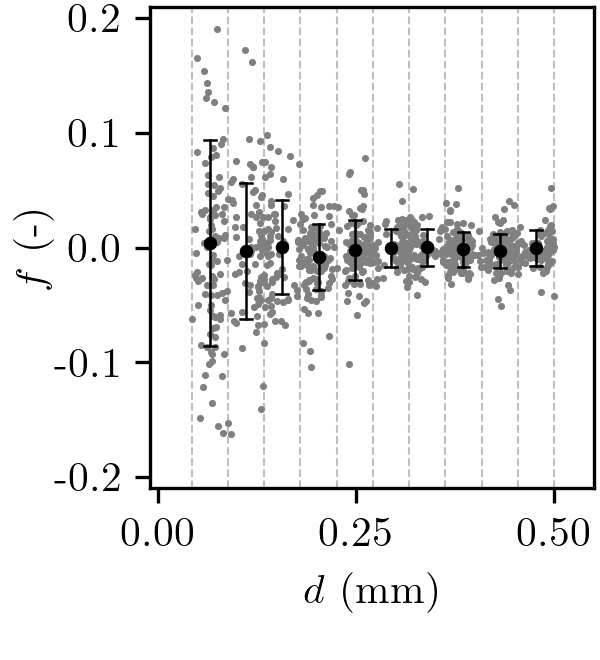}} 
    \caption{Scatter plot depicting the grain-averaged equivalent elastic strain error metric, $f$, for the simulations performed with mesh density $\bar{N}=1000$, as a function of their distance from the nearest control surface, $d$, at macroscopic strains of \subref{subfig:strain_error_bcs_domain_compare_scatter_1000_1} \SI{0.015}{\percent}, \subref{subfig:strain_error_bcs_domain_compare_scatter_1000_2} \SI{0.2}{\percent}, and \subref{subfig:strain_error_bcs_domain_compare_scatter_1000_3} \SI{10}{\percent}. Small gray dots indicate values of $f$ for individual grains, while large dark dots indicate the average values and error bars the standard deviations, both considering the data points within binned regions denoted by gray dashed lines.}
    \label{fig:strain_error_bcs_domain_compare_scatter_1000}
\end{figure}

We next turn to an investigation of the effects of boundary conditions on the observed behavior.  As mentioned previously, the mechanical response of the grains located at or close to the control surfaces (where constant velocities are imposed) may not be fully representative of the bulk behavior. This is why we previously used only the grains located in the central region of the specimen, which are well away from the control surfaces. Here, the goal is to quantify how ``deep'' inside a simulation volume this effect actually extends.  To do so, we run a simulation for which the (constant velocity) boundary conditions were applied directly to the central volume, as illustrated in Figure \ref{subfig:mesh_50_bcs_newbcs}.  Figure \ref{subfig:mesh_50_bcs} illustrates the corresponding tessellation. Compared to the nominal configuration, the grains located on the control surfaces are directly subjected to the imposed (constant) velocities. We then compare the mechanical response of all grains, in the reduced configuration, with those obtained in the nominal configuration (``ground truth'').  The question then is: how far from the control surfaces do grains need to be located so that their mechanical responses are not significantly biased by the boundary conditions? For the analysis, we consider the same 912 grains fully contained in the central volume as previously, as shown in Figure~\ref{subfig:mesh_50_bcs_reduced}.
To compare the mechanical response of the grains between the initial configuration and the new configuration, we define an error metric, $f$, akin to Equation~\ref{eq:strain_el_eq_grain}, though here comparing results between domains (quantified here via their axial length, $l$, of either 1 or 3) rather than between mesh densities:
\begin{equation}
    f_i\left(\bar{N}\right) = \frac{\bar{\varepsilon}_i^\text{e}(l=1,\bar{N}) - \bar{\varepsilon}_i^\text{e}(l=3,\bar{N})}{\bar{\varepsilon}_i^\text{e}(l=3,\bar{N})} \quad .
    \label{eq:strain_el_eq_grain_bcs}
\end{equation}
Interestingly, we find that the behavior is nearly constant as a function of mesh density ($\bar{N}) $ as well as macroscopic deformation state (we omit these figures for sake of brevity). We thus plot the error, $f$, as a function of the distance to the nearest control surface in the \qtyproduct[product-units = single]{1 x 1 x 1}{\milli\metre\cubed} domain. We present this data in Figure~\ref{fig:strain_error_bcs_domain_compare_scatter_1000} for the mesh density $\bar{N}=1000$ for sake of demonstration. We find that the largest differences between the domains occur in grains that reside near the control surfaces (i.e., where the boundary conditions are set), which diminishes as the distance from the control surfaces increases, converging to roughly steady state at a distance of roughly $d=\SI{0.3}{\milli\meter}$, which corresponds to a layer of three grains. We further note that there are still non-zero errors even at the highest values of $d=\SI{0.5}{\milli\meter}$, which can be explained by potential slight differences in grain morphology and mesh, non-zero simulation convergence criteria, as well as residual effects from the boundary conditions. Overall, we find that the clear display of high errors near the boundaries---coupled with the relative insensitivity to mesh density and macroscopic state---indicates that the boundary conditions have deleterious influence on the predicted behavior up to three grains from the control surface. In practice, this confirms the necessity of the use of a ``buffer layer'' between the region of interest and the control surfaces, in a simulation. One possibility to reduce the added computational cost of these buffer zones would be to utilize reduced-complexity models (such as J2 plasticity) to model the deformation inside the top and bottom region (see Figure~\ref{fig:minimal_bc}) instead of applying the boundary conditions directly to the central polycrystalline volume~\cite{Kapoor2019}.

\subsection{Effect of Elastic Anisotropy}
\label{sec:elas_aniso}

Here, in an objective to generalize the results to different materials, we investigate the influence of the degree of elastic anisotropy on the error in strain and orientation predictions. We begin by noting that simulations discussed prior to this point---utilizing modeling parameters associated with a copper alloy---have a Zener ratio, a metric quantifying the degree of elastic anisotropy which is defined as:
\begin{equation}
    \label{eq:zener}
    A = \frac{2C_{44}}{C_{11}-C_{12}},
\end{equation}
of approximately $A=3.21$. Here, we choose elastic constants to enforce significantly more elastically isotropic behavior. To this end, we alter the elastic parameters with a target of a significantly reduced Zener ratio. 

We choose a Zener ratio of $A=1.2$, as it is both nearly isotropic and representative of other structural materials (namely aluminum and its alloys). While we could choose modeling parameters to match those of a representative aluminum alloy to investigate a more elastically isotropic material, this may not facilitate the most direct comparison with the above results, as the bulk elastic behavior and the plastic behavior of the material will change. Consequently, to facilitate a more direct comparison with the above simulations, we enforce that the bulk elastic modulus of the material and the plasticity parameters remain fixed compared to previous simulations. As the bulk elastic modulus of a cubic crystal is dependent only on $C_{11}$ and $C_{12}$~\cite{hosford}, this leaves $C_{44}$ at our disposal for tuning the Zener ratio to the desired value of $A=1.2$. We present the elastic constants for the nearly elastically isotropic simulations in Table~\ref{tab:elastic_parms_iso}.
\begin{table}[H]
    \centering
    \label{tab:elastic_parms_iso}
    \caption{Single crystal elastic constants for FCC material with a Zener ratio of $A=1.2$.}
    \begin{tabular}{cccc}
        $C_{11}$ (\SI{}{\giga\pascal}) & $C_{12}$ (\SI{}{\giga\pascal}) & $C_{44}$ (\SI{}{\giga\pascal}) \\
        \hline
        168.4 & 121.4 & 28.2 \\
    \end{tabular}
\end{table}

\begin{figure}[htb!]
    \centering
    \subfigure[]{%
        \label{subfig:orientation_error_isotropic_metrics_1}
        \includegraphics[width=2in]{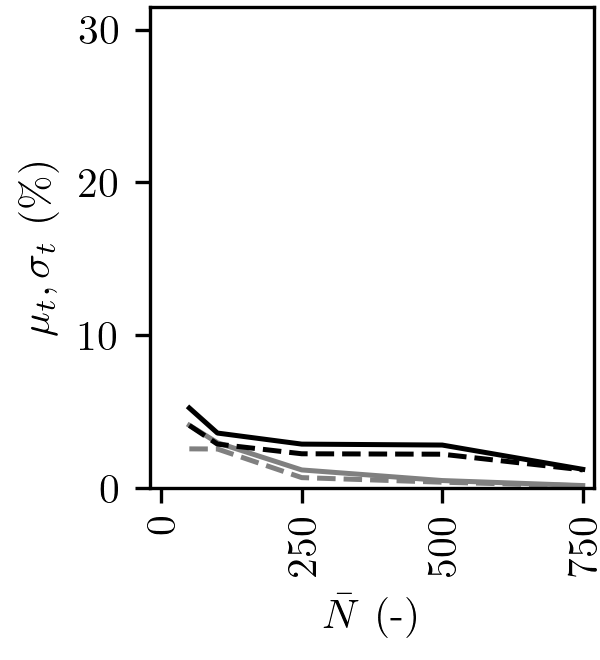}}
    \subfigure[]{%
        \label{subfig:orientation_error_isotropic_metrics_2}
         \includegraphics[width=2in]{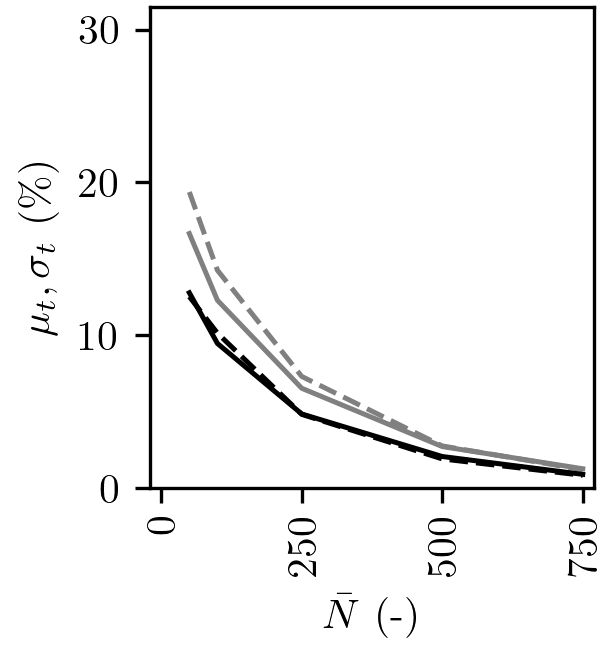}}
    \subfigure[]{%
        \label{subfig:orientation_error_isotropic_metrics_3}
        \includegraphics[width=2in]{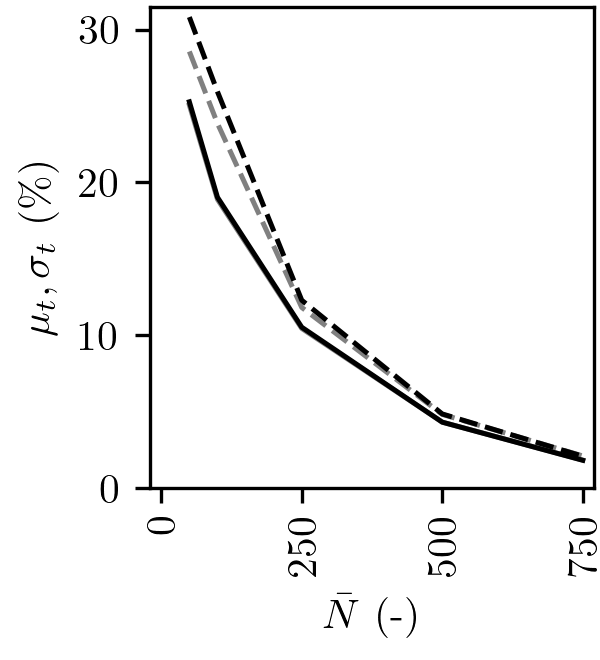}} 
    \caption{Mean ($\mu_t$, solid line) and standard deviation ($\sigma_t$, dashed line) of the grain-averaged orientation error metric, $t$, as a function of mesh density, $\bar{N}$, at macroscopic strains of \subref{subfig:orientation_error_isotropic_metrics_1} \SI{0.015}{\percent}, \subref{subfig:orientation_error_isotropic_metrics_2} \SI{0.2}{\percent}, and \subref{subfig:orientation_error_isotropic_metrics_3} \SI{10}{\percent}, for the simulations considering nearly isotropic elastic behavior. Black lines represent the statistics for the simulations considering nearly isotropic elastic behavior, while gray lines represent the statistics for the simulations considering anisotropic elastic behavior (i.e., a reproduction of the data in Figure~\ref{fig:strain_error_metrics}).}
    \label{fig:orientation_error_isotropic_metrics}
\end{figure}

We present the results below in Figures~\ref{fig:strain_error_isotropic_metrics} and~\ref{fig:orientation_error_isotropic_metrics} for the strain error metrics and orientation error metrics, respectively (compare to Figures~\ref{fig:strain_error_metrics} and~\ref{fig:orientation_error_metrics}, respectively, for the simulations performed with anisotropic elastic constants, and note that we omit the counterparts to Figures~\ref{fig:strain_error} and~\ref{fig:orientation_error} for sake of brevity). We note qualitatively similar trends as in the simulations performed with anisotropic elastic constants: errors are generally low in the elastic regime and higher in the plastic regime, and increase at all states as a function of a decrease in mesh density. The near-zero errors in the elastic regime (which---for a nearly isotropic material---we may expect mesh invariance) provides additional information regarding the errors present at these stages in the anisotropic simulations. Quantitatively, error values are on the same order of magnitude, albeit with slightly higher strain errors in the isotropic simulations versus the anisotropic simulations.
\begin{figure}[htb!]
    \centering
    \subfigure[]{%
        \label{subfig:strain_error_isotropic_metrics_1}
        \includegraphics[width=2in]{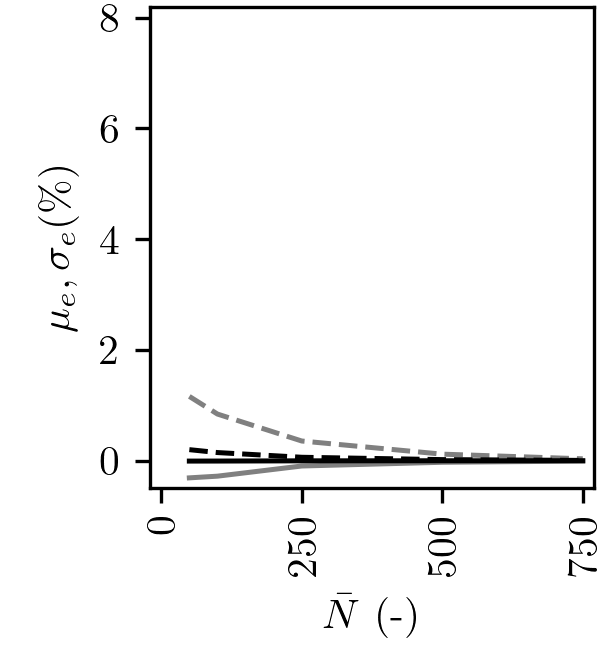}}
    \subfigure[]{%
        \label{subfig:strain_error_isotropic_metrics_2}
         \includegraphics[width=2in]{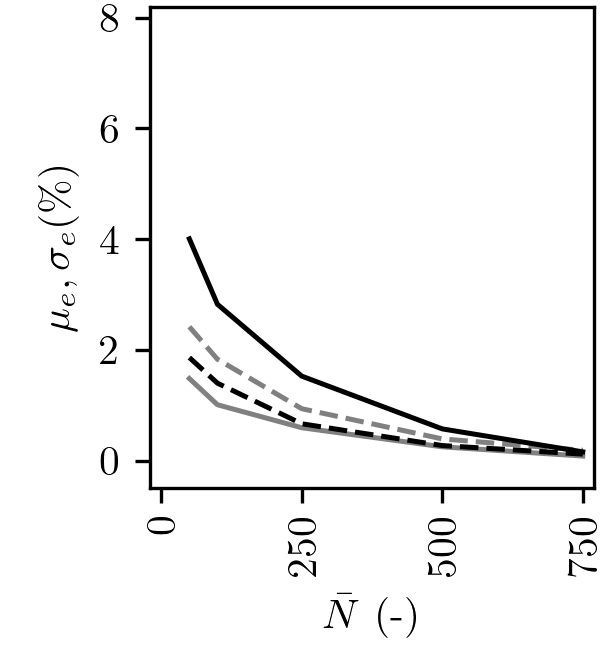}}
    \subfigure[]{%
        \label{subfig:strain_error_isotropic_metrics_3}
        \includegraphics[width=2in]{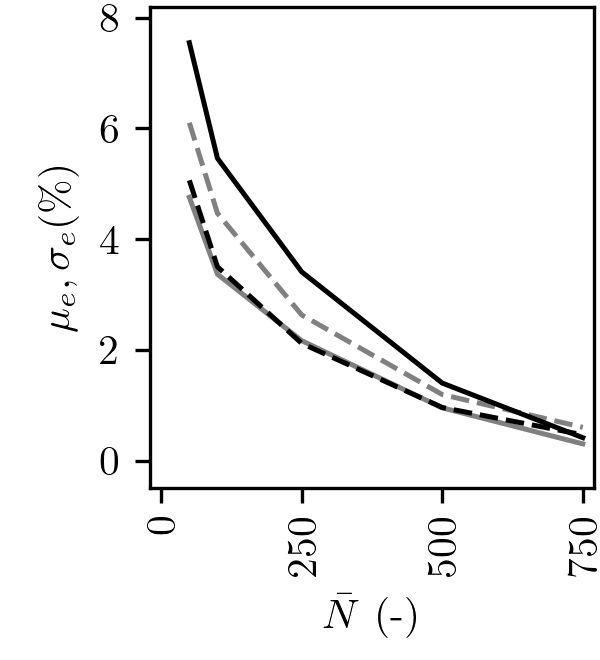}} 
    \caption{Mean ($\mu_e$, solid line) and standard deviation ($\sigma_e$, dashed line) of the grain-averaged equivalent elastic strain error metric, $e$, as a function of mesh density, $\bar{N}$, at macroscopic strains of \subref{subfig:strain_error_isotropic_metrics_1} \SI{0.015}{\percent}, \subref{subfig:strain_error_isotropic_metrics_2} \SI{0.2}{\percent}, and \subref{subfig:strain_error_isotropic_metrics_3} \SI{10}{\percent}, for the simulations considering nearly isotropic elastic behavior. Black lines represent the statistics for the simulations considering nearly isotropic elastic behavior, while gray lines represent the statistics for the simulations considering anisotropic elastic behavior (i.e., a reproduction of the data in Figure~\ref{fig:strain_error_metrics}).}
    \label{fig:strain_error_isotropic_metrics}
\end{figure}

\section{Conclusion}
\label{sec:conclusion}

We here present a study conducted to determine the sensitivity of grain-averaged deformation predictions from crystal plasticity finite element simulations to changes in mesh density. We find that error (both equivalent elastic strain and orientation) increases as mesh density decreases, owing to a lack of degree of freedoms necessary to capture true deformation fields inside grains. Further, we note that error (both equivalent elastic strain and orientation) is low in the elastic regime (even considering coarse finite element meshes), and is typically higher as plastic deformation initiates and progresses. Together, we find that a mesh density of 250 elements per grain is generally necessary in order to produce grain-averaged results generally within \SI{5}{\percent} of asymptotic behavior. However, even a simulation with a very coarse mesh ($\bar{N}=50$) will provide errors much lower than the variations of elastic strains between the grains of the polycrystal (by a factor of at least 4). Further, via simulations designed to explore the effects of boundary conditions, we find that a ``buffer layer'' of at least three grains is necessary to mitigate deleterious effects of boundary conditions on the predicted deformation response. This suggests the adoption of a paradigm in which larger far-field HEDM scans would be beneficial to provide indexed grains to construct buffer layers. We present these results to guide future experiment and simulation construction necessary to produce consistent results and mitigate deleterious effects due to boundary conditions.

\section*{Acknowledgments}

JL and MPK acknowledge financial support through the Department of Energy Nuclear Energy University Program award CFA-24-31412, as well as computational resources provided by the University of Alabama.

\section*{Data Availability}
The raw and processed data from this study are available upon reasonable request.

\bibliographystyle{elsarticle-num}
\bibliography{bibliography}

\end{document}